\documentclass[12pt,preprint]{aastex}
\usepackage{graphicx}
\usepackage{float}
\usepackage{multirow}
\usepackage{natbib}
\usepackage{appendix}
\usepackage{threeparttable}
\usepackage{amsmath,amssymb}

\slugcomment{KSUPT-14/2 \hspace{0.5truecm} January 2015}

\begin{document}

\title{Non-Gaussian Error Distribution of $\rm{^{7}Li}$ Abundance Measurements}

\author{Sara Crandall, Stephen Houston, and Bharat Ratra}

\affil{Department of Physics, Kansas State University, 
                 116 Cardwell Hall, Manhattan, KS 66506, USA} 
\email{sara1990@phys.ksu.edu, stephen8@phys.ksu.edu, and ratra@phys.ksu.edu}

\begin{abstract}
We construct the error distribution of $\rm{^{7}Li}$ abundance measurements for 66 observations (with error bars) used by \cite{Spite12} that give $\rm{A(Li)}=2.21 \pm 0.065$ (median and 1$\sigma$ symmetrized error). This error distribution is somewhat non-Gaussian, with larger probability in the tails than is predicted by a Gaussian distribution. The 95.4\% confidence limits are 3.0$\sigma$ in terms of the quoted errors. We fit the data to four commonly used distributions: Gaussian, Cauchy, Student's t, and double exponential with the center of the distribution found with both weighted mean and median statistics. It is reasonably well described by a widened $n=8$ Student's $t$ distribution. Assuming Gaussianity, the observed A(Li) is 6.5$\sigma$ away from that expected from standard Big Bang nucleosynthesis given the $Planck$ observations \citep{Coc2014}. Accounting for the non-Gaussianity of the observed A(Li) error distribution reduces the discrepancy to 4.9$\sigma$, which is still significant. 
\end{abstract}

\maketitle
\section{Introduction}
\label{Intro}

Standard Big Bang nucleosynthesis (BBN) is used to model the production of light elements D, $^3$He, $^4$He, and $^7$Li during the first 20 minutes of the development of the universe. With the baryon density determined by cosmic microwave background (CMB) observations, standard BBN predicts abundances of D, $^3$He, and $^4$He that are in good accord with the observed abundances, however, there is a large discrepancy for $^7$Li. The observed $^{7}\rm{Li}$ abundance appears to be depleted by a factor of about 3 when compared to the prediction of standard BBN in conjunction with CMB data. For recent reviews see \cite{Jedamzik2009}, \cite{Spite10}, \cite{Steigman2010}, \cite{Fields2011}, \cite{Frebel2011}, \cite{Spite12}, and \cite{Coc2014}.

$^7$Li is observed in the atmospheres of stars. It is best to observe old main-sequence and subgiant stars that are formed from primordial clouds because they are believed to preserve their lithium abundance. Very metal-poor stars are not sampled because they may not represent a lithium-rich star. $^7$Li is a very fragile isotope that is destroyed at temperatures above $2.5\times10^{6}$ K. In stars that are too hot, convection within the star occurs close to the surface and leads to a mixing of the atmosphere and the hotter, deeper layers of the star, and destruction of $^{7}\rm{Li}$, hence very hot stars are also not sampled. This has led to an emphasis on accurate modeling of the formation of Li. Ultimately, stellar $^{7}$Li is thought to be preserved and provides the best representation of the primordial abundance in warm, metal-poor dwarf, or turnoff, stars \citep{Spite12}. A(Li) determined from these stars is lower than expected. While some have argued for higher observed primordial $^{7}\rm{Li}$ abundances \citep[see e.g.,][]{Melendez2004,Melendez2010,Howk2012}, consistent with that expected from the CMB anisotropy data determination of the baryon density, here we focus on the more popular belief that standard BBN has a $^{7}\rm{Li}$ problem and try to more carefully quantify this discrepancy than has previously been done.\footnote{There are many discussions of possible mechanisms that might be responsible for this discrepancy. See, for instance \cite{Jedamzik2006}, \cite{Coc2007}, \cite{Chakraborty2011}, \cite{Fields2011}, \cite{Erkem2012}, \cite{Cyburt2012}, \cite{Ouyed2013}, and \cite{Kusakabe2014}.}

The conventional assumption is that data errors have a Gaussian distribution when the data is of good enough quality.\footnote{For instance, this is used when determining constraints from CMB anisotropy data \citep[see e.g.,][]{Ganga1997,Ratra1999,Chen2004,Bennett2013} and has been tested for such data \citep[see e.g.,][]{Park2001,Ade2013}.} Here we follow \cite{Chen2003a}\footnote{\cite{Chen2003a} examine the error distribution of 461 Hubble constant measurements from Huchra's list (many more than the 66 A(Li) measurements we will consider here). They discovered that the Hubble constant error distribution was very non-Gaussian and also confirmed the earlier \cite{Gott2001} median statistics estimate of the Hubble constant, more recently confirmed by \cite{Chen2011}, $H_{0}=68\pm2.8$ km s$^{-1}$ Mpc$^{-1}$. It is reassuring that many recent determinations of the Hubble constant agree with this median statistics estimate \citep[see e.g.,][]{Calabrese2012, Sievers2013, Holanda2014, Wang2014}.} and consider the A(Li) data in this context. We construct the distribution of the errors of the lithium abundance, ${\rm{A}_{i}\rm{(Li)}}_{-\sigma_{i}^{l}}^{+\sigma_{i}^{u}}$ (where $\sigma_{i}^{u(l)}$ is the 1 standard deviation upper (lower) error for the $i^{\rm{th}}$ abundance measurement), which is a plot of the number of measurements as a function of the number of standard deviations ($N_{\sigma}$) the measurement deviates from a central estimate A(Li)$_{\rm{CE}}$. Here 
\begin{equation}
\label{eq:1}
N_{\sigma_{i}}={\frac{{\rm{A(Li)}}_{i}-\rm{A(Li)}_{\mathrm{CE}}}{\sigma_{i}^{l}}}
\end{equation}
when ${\rm{A(Li)}}_{i}<\rm{A(Li)}_{CE}$ and
\begin{equation}
\label{eq:2}
N_{\sigma_{i}}={\frac{{\rm{A(Li)}}_{i}-\rm{A(Li)}_{\mathrm{CE}}}{\sigma_{i}^{u}}}
\end{equation}
when ${\rm{A(Li)}}_{i}>\rm{A(Li)}_{CE}$. In this paper $\rm{A(Li)_{CE}}$ is the central estimate determined using either the weighted mean or the median technique. We find that the error distribution has larger tails than predicted by a Gaussian distribution. \citet[Fig. 1]{Spite10}, have already noted the non-Gaussianity of the $^{7}\rm{Li}$ abundance error distribution. Here we present a quantitative analysis of this non-Gaussianity and fit the observed error distribution to various other commonly used distributions to determine which one provides a better fit. 

In Section \ref{DS} we summarize the criteria used to compile the \cite{Spite12} data. In this paper we are mostly concerned with the 66 (of 77) measurements that have error estimates. Section \ref{Analysis} describes our statistical analyses of the abundance data including finding central estimates using weighted mean and median statistics, analyzing the distribution of errors around the central estimates, and fitting several popular probability distribution functions to the observed error distribution. We conclude in Section \ref{Conclusion}.

\section{Data selection}
\label{DS}
To ensure that the A(Li) data  is representative of the primordial universe, \cite{Spite12}  select stars that have a metallicity that falls within the range $-2.8 \leq [\rm{Fe/H}] \leq -2.0$. If the metallicity is larger, the star is too young to be considered and if the metallicity is smaller, the star fails to lie on the lithium plateau that can be seen in \cite{Spite12}. Next, they retain only those stars with shallower convection resulting in atmospheres that are good indicators of the primordial abundance of lithium. This shallow convection is seen in warm metal-poor stars with an effective temperature $T_{\rm{eff}} \geq 5900$ K. These constraints result in a sample of 77 measurements. We do not consider two of the 77 measurements because they are upper bounds. In most of our statistical analyses we consider only 66 values of the remaining 75 (with an $\rm{A(Li)}=2.20 \pm 0.064$ (mean and 1$\sigma$ error)) measurements that have error bars.\footnote{The remaining nine measurements have a quoted error of $\sigma=0.01$, but this accounts only for that from the signal to noise ratio, and is not the full error, so we do not include these in our analyses here.} These values are listed in Table \ref{table:Lithium}.\footnote{The errors quoted in these references were found by adding in quadrature errors from stellar parameters and those from equivalent widths. \cite{Bon07} argue that the effective temperature error is larger than previously thought, and dominates the overall error, resulting in a constant error for all measurements. Given that we find the A(Li) error distribution is a non-Gaussian, to illustrate the possibility that this is the result of unaccounted-for systematic error, in the Appendix we repeat our analysis for the case with a constant error for all measurements. We thank F. Spite for helpful discussion on this matter.} For these 66 measurements $\rm{A(Li)}=2.21 \pm 0.065$ (median and 1$\sigma$ symmetrized error).

\begin{table*}
\centering
\caption{66 Lithium Abundance Measurements from \cite{Spite12}.}
\tiny
\vspace{2mm}
\begin{tabular}{l c c c c}
\hline\hline 
{Star Number/Name}& {A(Li)\tablenotemark{a}}& {$\sigma^{l}$\tablenotemark{b}}&{$\sigma^{u}$\tablenotemark{c}}&{Reference}\\
\hline
{47480} & {2.17} & {0.061} & {0.061} & \multirow{23}{*}{\cite{Char05}}\\
{23344} & {2.13} & {0.026} & {0.026} &{}\\
{36513} & {2.15} & {0.094} & {0.078} & {}\\
{61361} & {2.25} & {0.063} & {0.027} & {}\\
{106468} & {2.24} & {0.025} & {0.024} & {}\\
{65206} & {2.07} & {0.057} & {0.048} & {}\\
{87467} & {2.25} & {0.067} & {0.057} & {}\\
{34630} & {2.12} & {0.052} & {0.049} & {}\\
{72461} & {2.22} & {0.066} & {0.059} & {}\\
{8572} & {2.24} & {0.027} & {0.026} & {}\\
{96115} & {2.22} & {0.028} & {0.021} & {}\\
{88827} & {2.31} & {0.052} & {0.048} & {}\\
{68321} & {2.15} & {0.033} & {0.031} & {}\\
{48152} & {2.25} & {0.029} & {0.027} & {}\\
{12529} & {2.22} & {0.024} & {0.046} & {}\\
{61545} & {2.15} & {0.075} & 0.062 & {}\\
{59109} & {2.27} & {0.057} & {0.059} & {}\\
{59376} & {2.11} & {0.019} & {0.020} & {}\\
{36430} & {2.35} & {0.042} & {0.040} & {}\\
{87693} & {2.20} & {0.041} & {0.045} & {}\\
{111372} & {2.28} & {0.086} & {0.096} & {}\\
{102337} & {2.30} & {0.062} & 0.053 & {}\\
{115704} & {2.15} & {0.038} & 0.041 & {}\\
\hline
{LP0815-0043} & {2.20} & {0.035} & {0.035} & \multirow{19}{*}{\cite{Melendez2010}}\\
{BD-13 3442} & {2.18} & {0.035} & {0.035} &{}\\
{BD+03 0740} & {2.17} & {0.035} & {0.035} & {}\\
{BD+09 2190} & {2.13} & {0.035} & {0.035} & {}\\
{BD+24 1676} & {2.21} & {0.035} & {0.035} & {}\\
{LP0635-0014} & {2.28} & {0.035} & {0.035} & {}\\
{CD-35 14849} & {2.29} & {0.035} & {0.035} & {}\\
{BD-10 0388	} & {2.21} & {0.035} & {0.035} & {}\\
{BD-04 3208} & {2.30} & {0.035} & {0.035} & {}\\
{HD 338529} & {2.23} & {0.035} & {0.035} & {}\\
{BD+02 3375} & {2.21} & {0.035} & {0.035} & {}\\
{HD 084937} & {2.26} & {0.035} & {0.035} & {}\\
{G011-044} & {2.30} & {0.035} & {0.035} & {}\\
{HD 24289} & {2.24} & {0.035} & {0.035} & {}\\
{BD+34 2476} & {2.23} & {0.035} & {0.035} & {}\\
{BD+42 3607} & {2.22} & {0.035} & {0.035}& {}\\
{BD+09 0352} & {2.21} & {0.035} & {0.035} & {}\\
{HD 19445} & {2.22} & {0.035} & {0.035} & {}\\
{HD 74000} & {2.20} & {0.035} & {0.035} & {}\\
\hline
{+26 3578} & {2.28} & {0.070} & {0.070} & \multirow{2}{*}{\cite{Aoki09}}\\
{042-003} & {2.26} & {0.070} & {0.070} & {}\\
\hline
{BD+03$^{\circ}$ 740} & {2.13} & {0.074} & {0.074} & \multirow{13}{*}{\cite{Hosford09}}\\
{BD+09$^{\circ}$ 2190} & {2.10} & {0.084} & {0.084} & {}\\
{BD-13$^{\circ}$ 3442} & {2.15} & {0.057} & {0.057} & {}\\
{BD+26$^{\circ}$ 2621} & {2.17} & {0.070} & {0.070} & {}\\
{BD+20$^{\circ}$ 2030} & {2.07} & {0.068} & {0.068} & {}\\
{LP815-43} & {2.17} & {0.070} & {0.070} & {}\\
{BD+24$^{\circ}$ 1676} & {2.16} & {0.009} & {0.009} & {}\\
{LP635-14} & {2.12} & {0.074} & {0.074} & {}\\
{CD-71$^{\circ}$ 1234} & {2.20} & {0.035} & {0.035} & {}\\
{BD+26$^{\circ}$ 3578} & {2.15} & {0.053} & {0.053} & {}\\
{CD-35$^{\circ}$ 14849} & {2.24} & {0.025} & {0.025} & {}\\
{HD84937} & {2.17} & {0.066} & {0.066} & {}\\
{HD74000} & {2.05} & {0.083} & {0.083} & {}\\
\hline
{CS29518-020} & {2.13} & {0.090} & {0.090} & \multirow{4}{*}{\cite{Bon07}}\\
{CS30301-024} & {2.10} & {0.090} & {0.090} & {}\\
{CS29499-060} & {2.16} & {0.090} & {0.090} & {}\\
{CS31061-032} & {2.22} & {0.090} & {0.090} & {}\\
\hline
{BS17572-100} & {2.17} & {0.090} & {0.090} & \multirow{3}{*}{\cite{Sbordone2010}}\\
{CS22950-173} & {2.23} & {0.090} & {0.090} & {}\\
{CS29514-007} & {2.24} & {0.090} & {0.090} & {}\\
\hline
{G37-37} & {2.24} & {0.120} & {0.120} & \multirow{2}{*}{\cite{King12}}\\
{G130-65} & {2.30} & {0.120} & {0.120} & {}\\
\hline
\end{tabular}
\tablenotetext{a}{Following the advice of M. Spite, for stars with both a main sequence and a sub-giant branch lithium abundance measurement, we list the average of the two values.} 
\tablenotetext{b}{$\rm{L}$ower 1$\sigma$ error.}
\tablenotetext{c}{Upper 1$\sigma$ error.}
\label{table:Lithium}
\end{table*}

\section{Analysis}
\label{Analysis}
\subsection{Weighted mean and median statistics}
\label{WM,Med}
We first compute central estimates using two different statistical techniques: weighted mean and median statistics. 

The conventional weighted mean statistics has the advantage that a goodness-of-fit criterion can be obtained. The standard weighted mean formula \citep{Podariu01} is
\begin{equation}
{\rm{A(Li)}}_{\mathrm{wm}}={\frac{\sum\limits_{i=1}^{N}{\rm{A(Li)}}_{i}/\sigma_{i}^2}{\sum\limits_{i=1}^{N}1/\sigma_{i}^2}},
\end{equation}
where $\sigma_{i}$ is the one standard deviation error of $i=1,2,....,N$ measurements.\footnote{For measurements that have different upper and lower errors, we use the average of the two errors in the weighted mean formula.} The weighted standard deviation is
\begin{equation}
\sigma_{\mathrm{wm}}=\left(\sum_{i=1}^{N}1/\sigma_{i}^{2}\right)^{-1/2}.
\end{equation}
We can also determine a goodness of fit $\chi^{2}$ by
\begin{equation}
\chi^{2}=\frac{1}{N-1}\sum_{i=1}^{N}\frac{{\rm{(A(Li)}}_{i}-\rm{A(Li)}_{\mathrm{wm}})^{2}}{\sigma_{i}^{2}}.
\end{equation}
The number of standard deviations that $\chi$ deviates from unity is a measure of good-fit and is given by
\begin{equation}
N_{\sigma}=|\chi-1|\sqrt{2(N-1)}.
\end{equation}
Here $1/\sqrt{2(N-1)}$ is the expected error of $\chi$. Hence $N_{\sigma}$ represents the number of standard deviations $\chi$ deviates from unity. A large $N_{\sigma}$ can be caused by correlations between measurements, systematic error, or invalidity of the Gaussian assumption.

The second statistical tool we use in analyzing the $^{7}\rm{Li}$ measurements is median statistics. For a detailed description of median statistics see \cite{Gott2001}.\footnote{For other applications see \cite{Chen2003b}, \cite{Hodge2008}, \cite{Mamajek2008}, \cite{Bourne2011}, \cite{Shafieloo2011}, \cite{Croft2011}, \cite{Andreon2012}, \cite{Farooq2013}, \cite{Crandall2014}, and references therein.} In using this method we assume that the measurements are statistically independent, and have no systematic error as a whole. As the number of measurements increase to infinity, the median will become a true value. This method has the advantage that it does not make use of the quoted errors. Consequentially this will result in a larger uncertainty than that of the weighted mean method.

If the measurements are statistically independent then there is a 50\% chance of finding any value above or below the median. As described in \cite{Gott2001}, the probability of the $n^{\rm{th}}$ measurement being higher or lower than the true median is
\begin{equation}
P_{n}=\frac{2^{-N}N!}{n!(N-n)!}.
\end{equation}

With the use of weighted mean statistics we find a central estimate of $\rm{A(Li)}_{\mathrm{wm}}=2.20$ with a weighted error of $\sigma_{\mathrm{wm}}=4.43 \times 10^{-3}$. We also find $\chi^{2}=2.41$ and $N_{\sigma}=6.28$; such a large value is cause for concern. Using median statistics we find a lithium abundance central estimate of $\rm{A(Li)}_{\mathrm{med}}=2.21$ with a 1$\sigma$ range of $2.13\leq \rm{A(Li)}_{\mathrm{med}} \leq2.26$, and a 2$\sigma$ range of $2.05\leq \rm{A(Li)}_{\mathrm{med}} \leq2.31$.\footnote{We also use median statistics to analyze the original set of 75 abundance values of \cite{Spite12} since this method does not depend on the quoted errors. This results in a central estimate of  $\rm{A(Li)}_{\mathrm{med}}=2.21$ with a 1$\sigma$ range of $2.13\leq \rm{A(Li)}_{\mathrm{med}} \leq2.26$, and a 2$\sigma$ range of $2.07\leq \rm{A(Li)}_{\mathrm{med}} \leq2.31$, reassuringly consistent with the values determined above.}

Our central estimates are very similar to that of \cite{Spite12}. There they use a straight mean to estimate of $\rm{A(Li)}=2.20 \pm 0.064$ for the set of 75 measurements.

\subsection{Error distribution and distribution functions}
\label{DF}
\begin{center}
\begin{figure}[H]
\advance\leftskip-1.25cm
\advance\rightskip-1.25cm
\includegraphics[height=68mm,width=95mm]{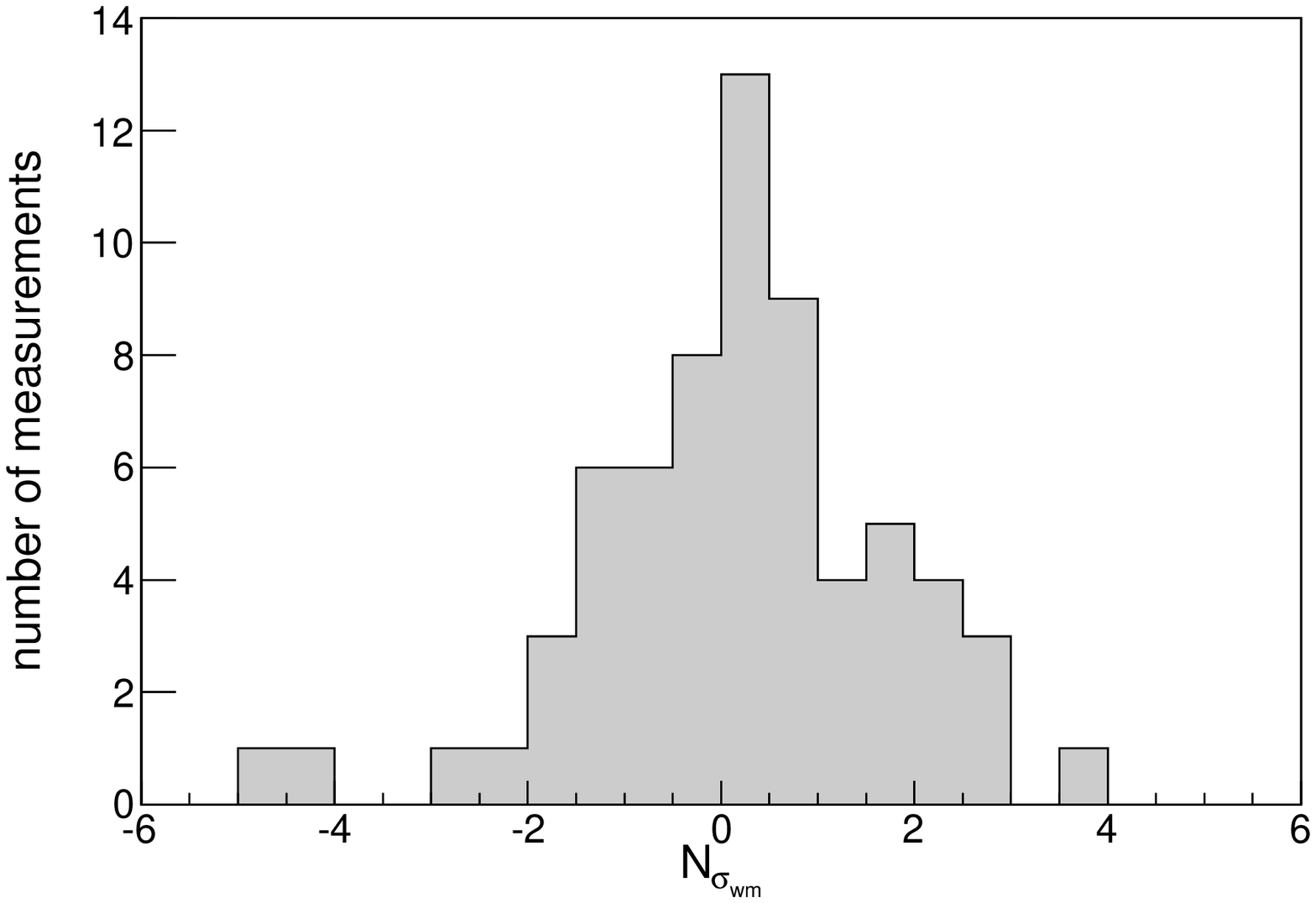}
\includegraphics[height=68mm,width=95mm]{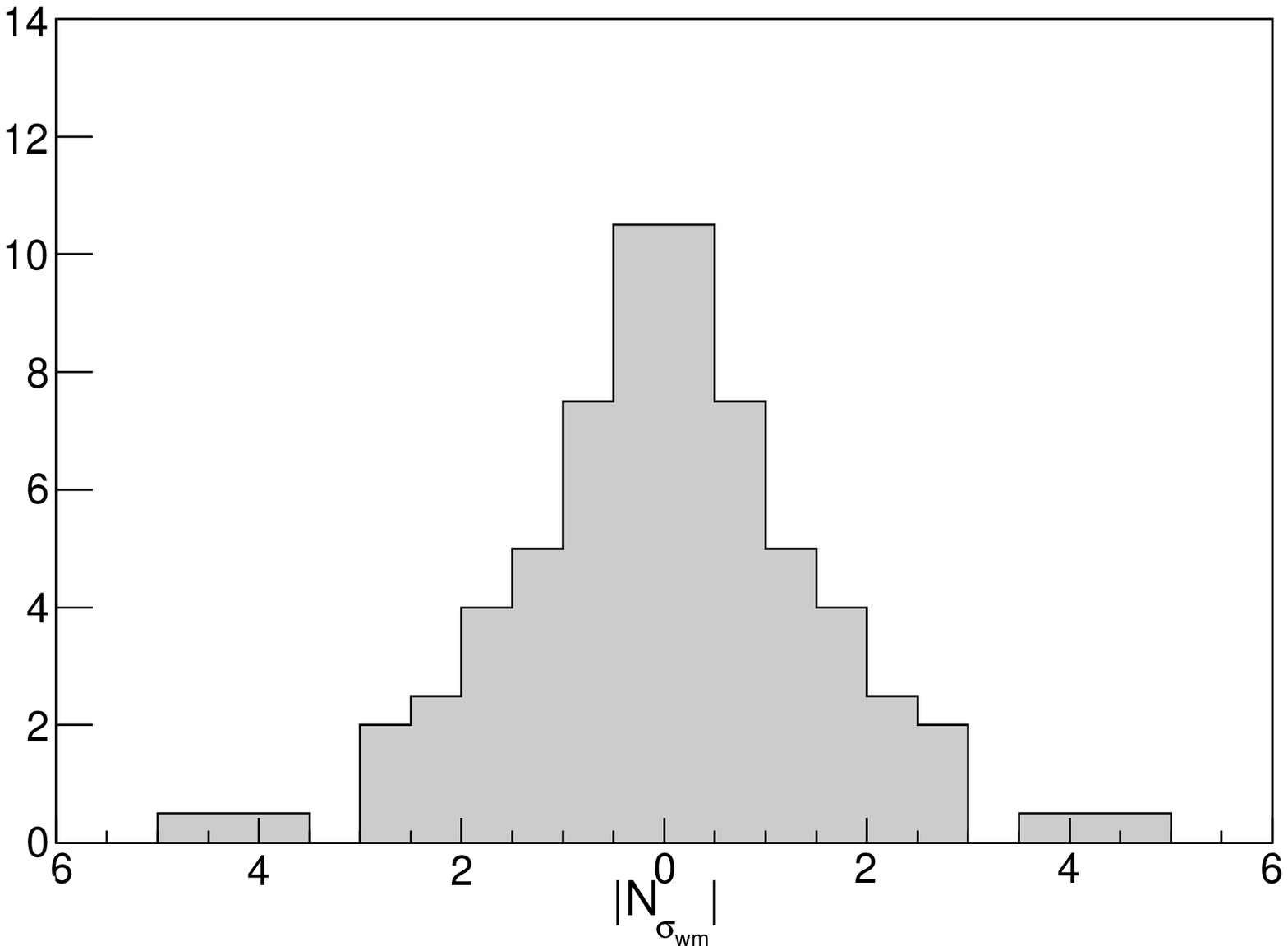}
\includegraphics[height=68mm,width=95mm]{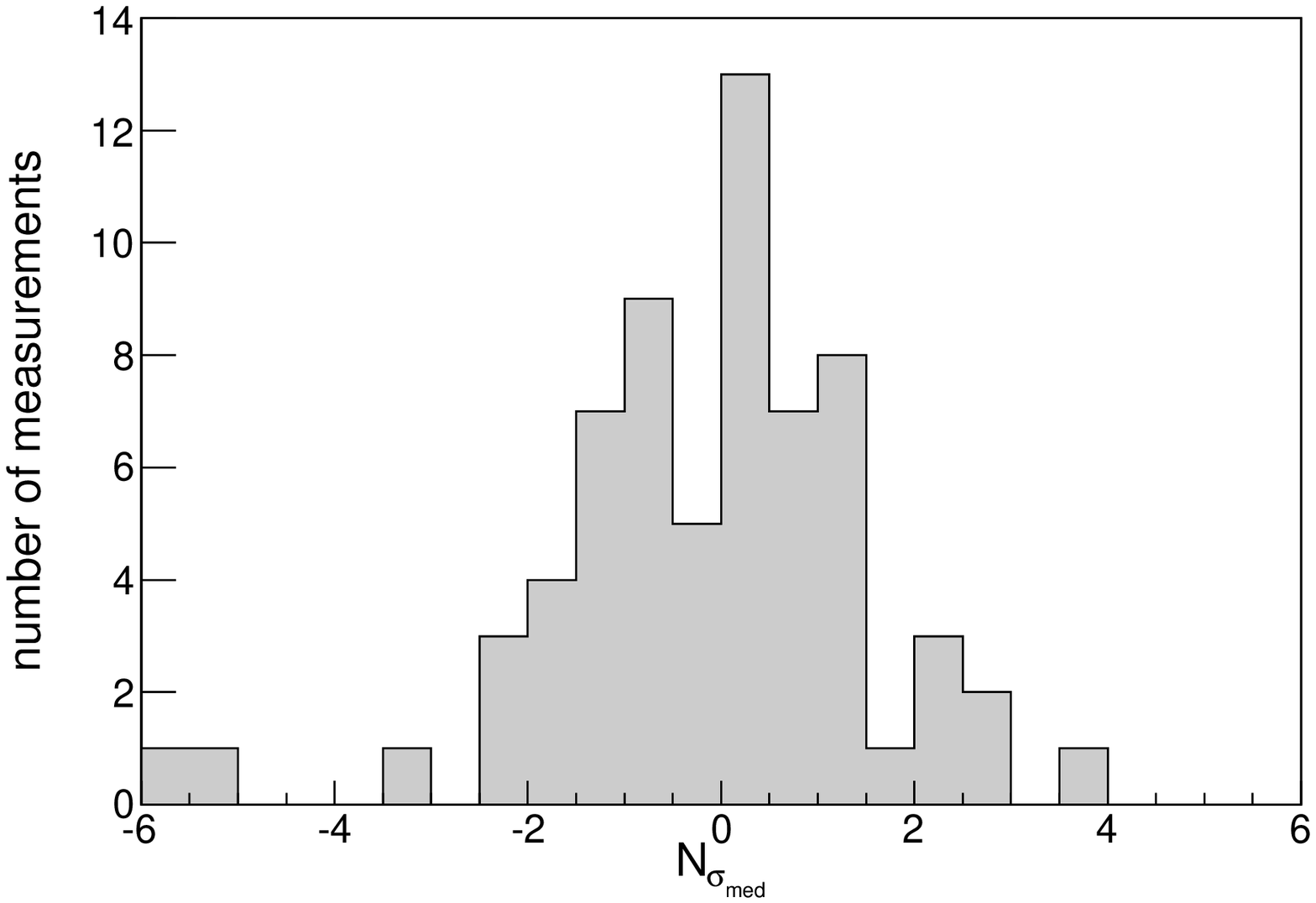}
\includegraphics[height=68mm,width=95mm]{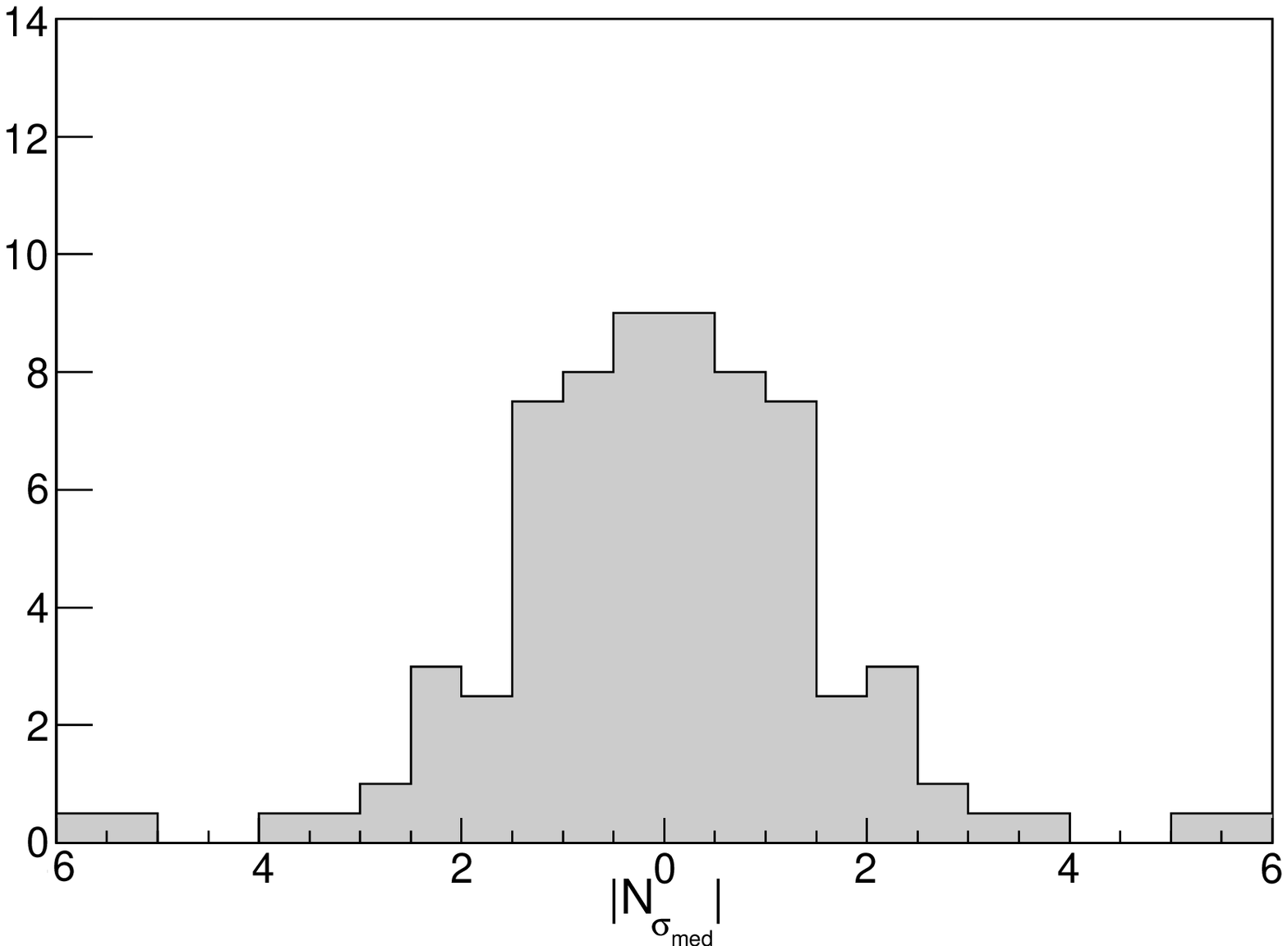}
\caption{Histograms of the error distribution in half standard deviation bins. The top (bottom) row uses the weighted mean (median) of the 66 measurements as the central estimate. The left (right) column shows the signed (absolute) deviation. In the left column plots, positive (negative) $N_{\sigma}$ represent a value that is greater (less) than the central estimate.} 
\label{figure:Nsig}
\end{figure}
\end{center}

It is of interest to determine the error distribution of the $^{7}\rm{Li}$ measurements. To do so, we plot the number of standard deviations each measurement deviates from the central estimate, as described in \cite{Chen2003a}. The formulae we use are in Eqs. ({\ref{eq:1}}) \& ({\ref{eq:2}}) above. We use both the weighted mean and median statistics A(Li)$_{\rm{CE}}$ central estimates. Figure \ref{figure:Nsig} shows the $N_{\sigma}$ and $|N_{\sigma}|$ histograms using both central estimates.\footnote{The un-binned data used to derive the signed error distribution in the left column of Fig. \ref{figure:Nsig} have a mean of $N_{\sigma}=0.19$ ($-0.14$), median of $N_{\sigma}=0.37$ (0.0), standard deviation $\sigma=1.56$ (1.61), skewness $-0.37 (-0.78)$ and a kurtosis of $0.61 (1.84)$ for the weighted mean (median) case.} 

For the weighted mean case, 68.3\% of the signed error distribution falls within $-1.88\leq N_{\sigma}\leq1.15$ while 95.4\% lies in the range of $-4.53\leq N_{\sigma}\leq2.03$ and the absolute magnitude of the error distribution have corresponding limits of $|N_{\sigma}|\leq 1.41$ and $|N_{\sigma}|\leq 3.0$ respectively. For the median statistics central estimate, 68.3\% of the signed error distribution falls within $-1.75\leq N_{\sigma}\leq1.15$ while 95.4\% lies within $-5.56\leq N_{\sigma}\leq2.25$ and the corresponding absolute magnitude limits are $|N_{\sigma}|\leq 1.31$ and $|N_{\sigma}|\leq 3.10$ respectively. Alternatively, when looking at the fraction of the data that falls within the $|N_{\sigma}| = 1$ and $2$ ranges respectively, we obtain $54.5\%$ and $81.8\%$ for the weighted mean case and $51.5\%$ and $81.8\%$ for the median one. Although these fractions are large, they are not as large as for a Gaussian distribution. The error distribution of the 66 $^{7}$Li abundance measurements used by \cite{Spite12} is somewhat non-Gaussian. It is of interest to determine if a well-known non-Gaussian distribution function provide a better fit to the data.

To do this, we follow \cite{Chen2003a} and bin our data so as to maximize the number of bins as well as the number of data points within each bin. This is best done by using the number of points per bin closest to the square root of the total number of measurements. We thus use 8 bins that are labeled by the integer $j=1,2,....,8$, with each bin width allowed to vary to maintain about 8 measurements per bin. Adjusting the bin width, $\Delta|N_{\sigma}|_{j}$, ensures equal probability in each bin\footnote{From this point forward, we focus only on the symmetric absolute error distribution.} for the assumed distribution function. Therefore, for any of the assumed distributions, $P(|N_{\sigma}|)$, there are 8 bins that are expected to contain 8.25 data points.

We use the average of each bin to represent the bin as a whole and estimate a goodness of fit from a $\chi^2$ analysis using
\begin{equation}
\chi^{2}=\sum_{j=1}^{8}\frac{[M(|N_{\sigma}|_j)-NP(|N_{\sigma}|_j)]^2}{NP(|N_{\sigma}|_j)}.
\end{equation} 
Here $M(|N_{\sigma}|)$ is the observed number of measurements in each bin and $N=66$ is the total number of measurements. We use $P(|N_{\sigma}|)$, the assumed probability distribution function, to compute the expected number of measurements in bin $j$, $NP(|N_{\sigma}|_{j})$. We continue on to calculate the reduced $\chi^2$, $\chi_{\nu}^2=\chi^2/\nu$, where $\nu$ is the number of degrees of freedom (the number of bins, 8, minus the number of fitting parameters and the number of constraints). From $\chi_{\nu}^2$ and $\nu$ we are able to compute how well the probability distribution describes the spread of the measurements.\footnote{We assume the bins are uncorrelated in this computation of the probability, which is not necessarily true. Therefore, it is best to view the probability determined from $\chi_{\nu}^2$ as simply a qualitative indicator of goodness of fit and place a more quantitative emphasis on the value of $\chi_{\nu}^2$.} The values of $\chi_{\nu}^2$ and the corresponding probability for the probability distribution function for both the weighted mean and median analyses are compiled in Table \ref{table:Goodness-of-Fit}.

\begin{deluxetable}{lcccccccc}
\tablecaption{Goodness-of-Fit}
\tablewidth{0pt}
\tabletypesize{\small}

\tablehead{
&\multicolumn{4}{c}{Weighted Mean}&\multicolumn{4}{c}{Median}\\
\hline
\colhead{Function\tablenotemark}& 
\colhead{Scale\tablenotemark{a}}& 
\colhead{$\chi_{\rm \nu}^{2{b}}$}&
\colhead{$\nu$\tablenotemark{b}} &
\colhead{Probability(\%)\tablenotemark{c}}&
\colhead{Scale\tablenotemark{a}}& 
\colhead{$\chi_{\rm \nu}^{2{b}}$}&
\colhead{$\nu$\tablenotemark{b}} &
\colhead{Probability(\%)\tablenotemark{c}}\\
\noalign{\vskip -6mm}
}
\startdata
\noalign{\vskip -0mm}
\noalign{\vskip 0mm}
Gaussian............................&	1&	65.8&	7&	   $<0.1$&		1&	  101&	7&	$<0.1$\\
\noalign{\vskip 2mm}
Gaussian............................&	1.7&	10.9&	6&	   $<0.1$&		1.8&   11.5&	6&	$<0.1$\\
\noalign{\vskip 2mm}
Cauchy...............................&	1&	6.70&	7&	   $<0.1$&		1&	 7.31&	7&	$<0.1$\\
\noalign{\vskip 2mm}
Cauchy...............................&	1.6&	5.54&	6&	   $<0.1$&		1.6&	  6.00&	6&	$<0.1$\\
\noalign{\vskip 2mm}
$n = 8$ Student's $t^{\rm{d}}$...............&	1&	25.3&	6&	   $<0.1$&		1&	  30.4&	6&	$<0.1$\\
\noalign{\vskip 2mm}
$n = 8$ Student's $t$...............&	2.6&	2.04&	5&	   6.9&		2.8&	 2.16&	5&	5.5\\
\noalign{\vskip 2mm}
Double Exponential...........&	1&	10.5&	7&	   $<0.1$&		1&	12.4&	7&	$<0.1$\\
\noalign{\vskip 2mm}
Double Exponential...........&	1.4&	8.76&	6&	   $<0.1$&		1.5&	  9.94&	6&	$<0.1$\\
\enddata
\tablenotetext{a}{For a Gaussian distribution, $N_{\sigma}=1$ corresponds to 1 standard deviation when the scale factor $S=1$. For the other cases, the scale factor varies with the width of the distribution to allow $\chi^{2}$ to be minimized.}
\tablenotetext{b}{$\chi_{\rm \nu}^{2}$ is the $\chi^{2}$ divided by the number of degrees of freedom $\nu$.}
\tablenotetext{c}{The probability that a random sample of data points drawn from the assumed distribution yields a value of $\chi_{\rm \nu}^{2}$ greater than or equal to the observed value for $\nu$ degrees of freedom. This probability assumes that the bins are uncorrelated, which is not necessarily true. Therefore, the probabilities should only be viewed as qualitative indicators of goodness of fit.}
\tablenotetext{d}{We find that for the Student's $t$ distribution, the $n=2$ and $S=1$ case gives a smaller reduced $\chi_{\nu}^{2}=8.25$ (8.89) with a probability of $<0.1$ ($<0.1$) for the weighted mean (median) case. However, when allowing the scale factor $S$ to vary, the $n=8$ case has a lower reduced $\chi_{\nu}^{2}$ than the $n=2$ case (also see footnote 11).} 
\label{table:Goodness-of-Fit}
\end{deluxetable}

Throughout this analysis we ensure that the constraint of the sum of the measurements being 66 is always satisfied. The degrees of freedom, $\nu$, is equal to 7 (since there are 8 bins under consideration) when considering no additional free parameters. We fit the data to four probability distribution functions with each distribution centered at $|N_{\sigma}|=0$ and only the absolute magnitude of the error distribution is used. In each case we also consider a corresponding probability distribution with a scale factor, $S$, that is used to vary the width of the distribution (consequently removing another degree of freedom) while minimizing $\chi^2$.

\begin{center}
\begin{figure}[H]
\advance\leftskip-1.25cm
\advance\rightskip-1.25cm
\includegraphics[height=68mm,width=95mm]{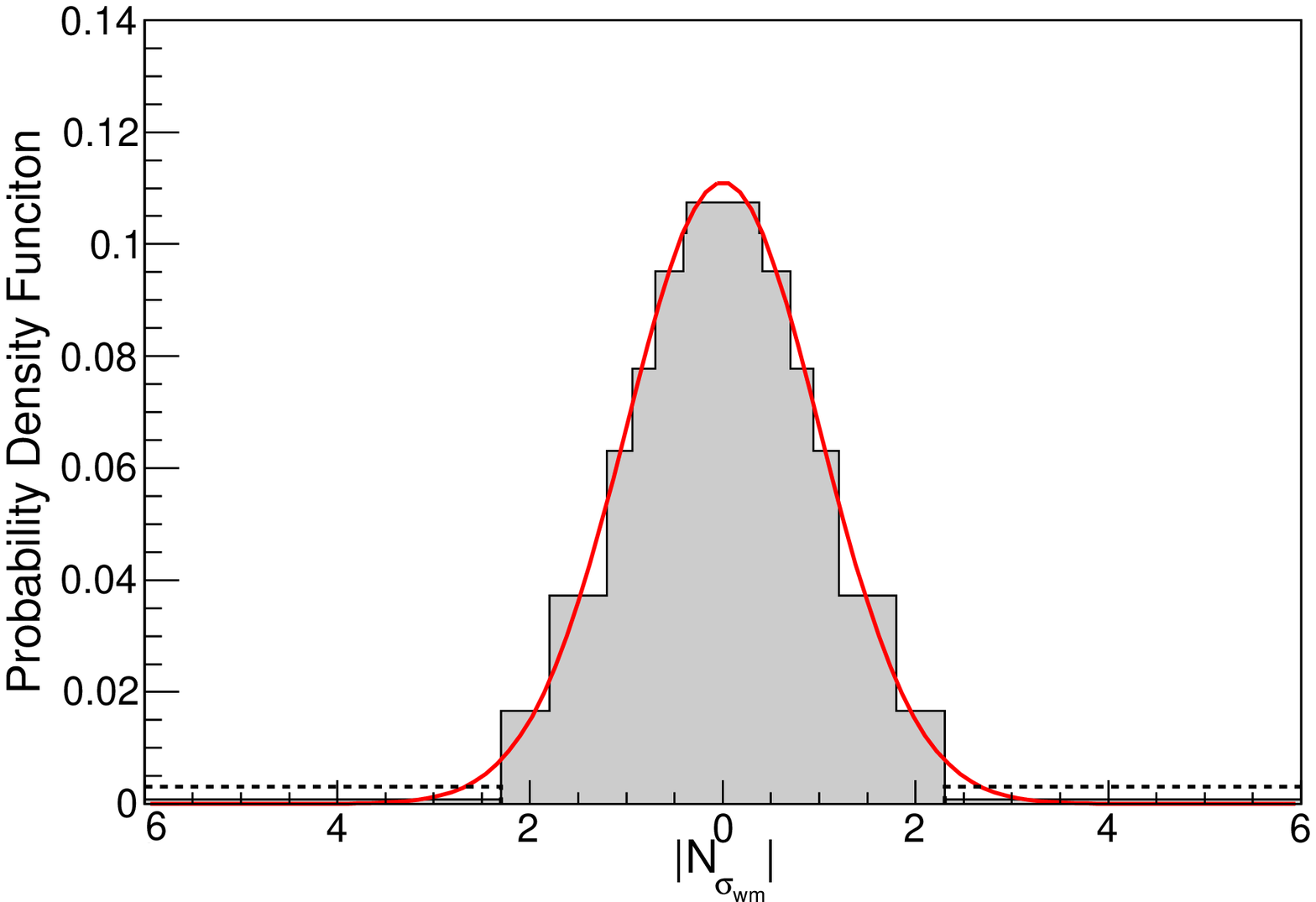}
\includegraphics[height=68mm,width=95mm]{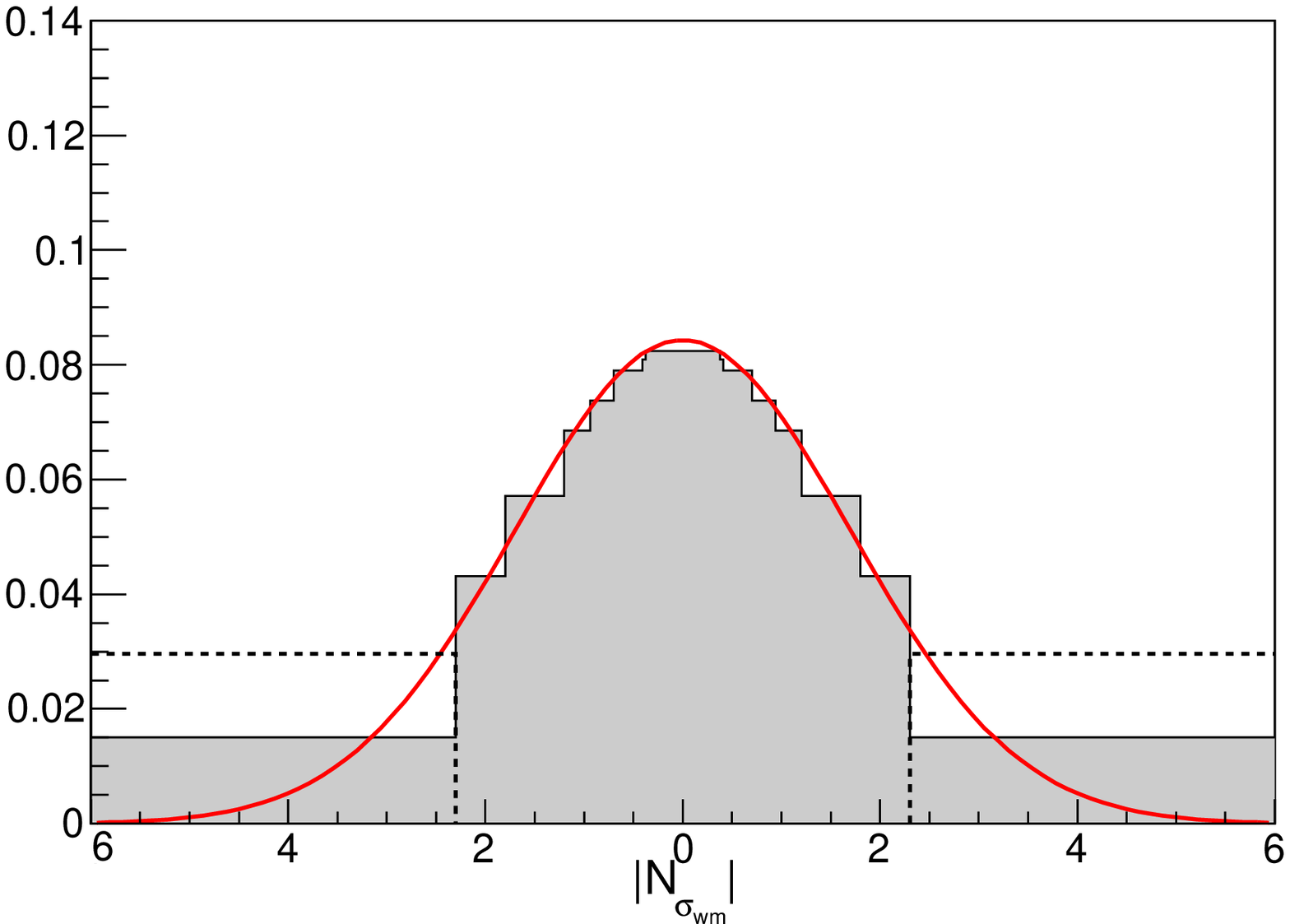}
\includegraphics[height=68mm,width=95mm]{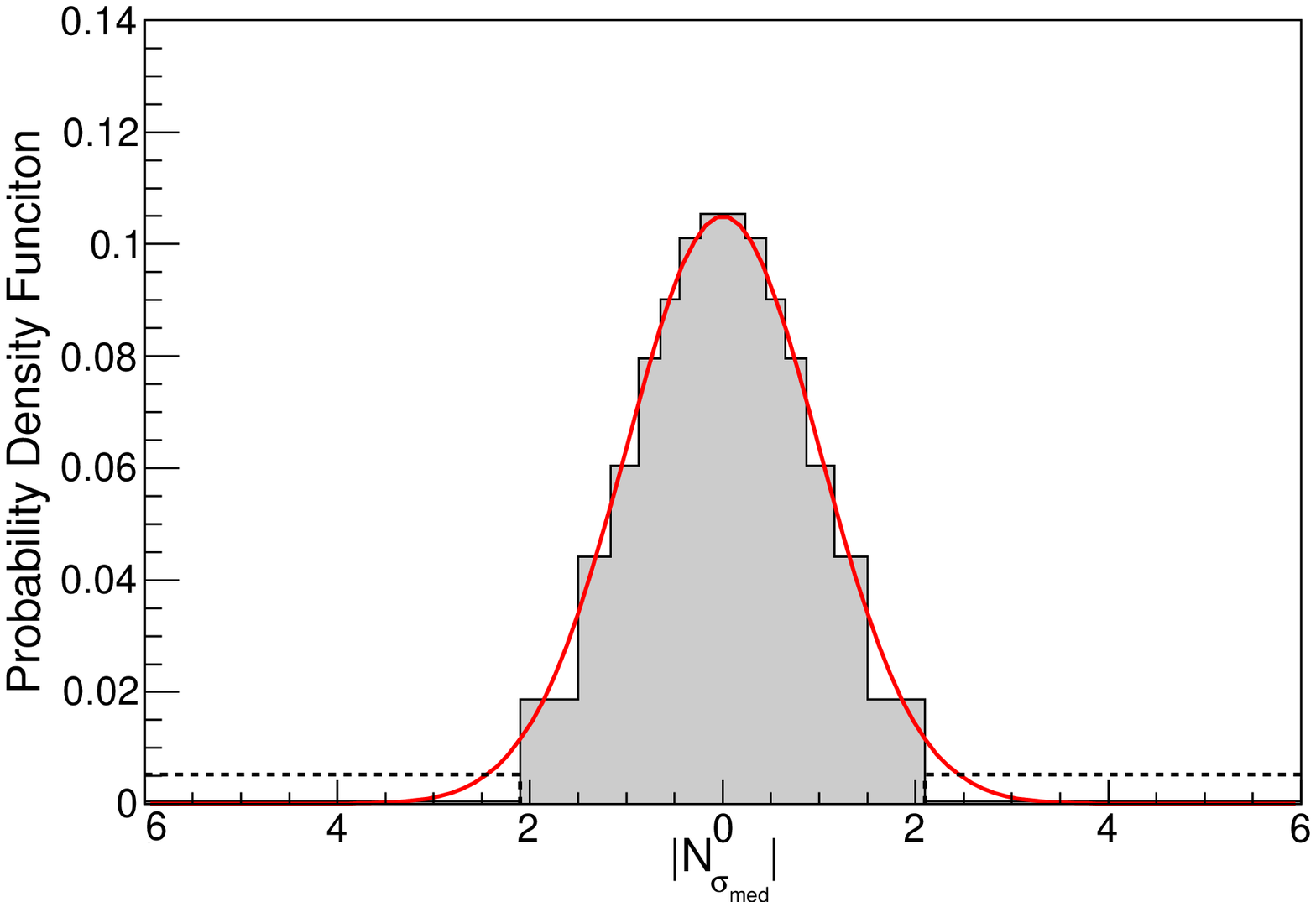}
\includegraphics[height=68mm,width=95mm]{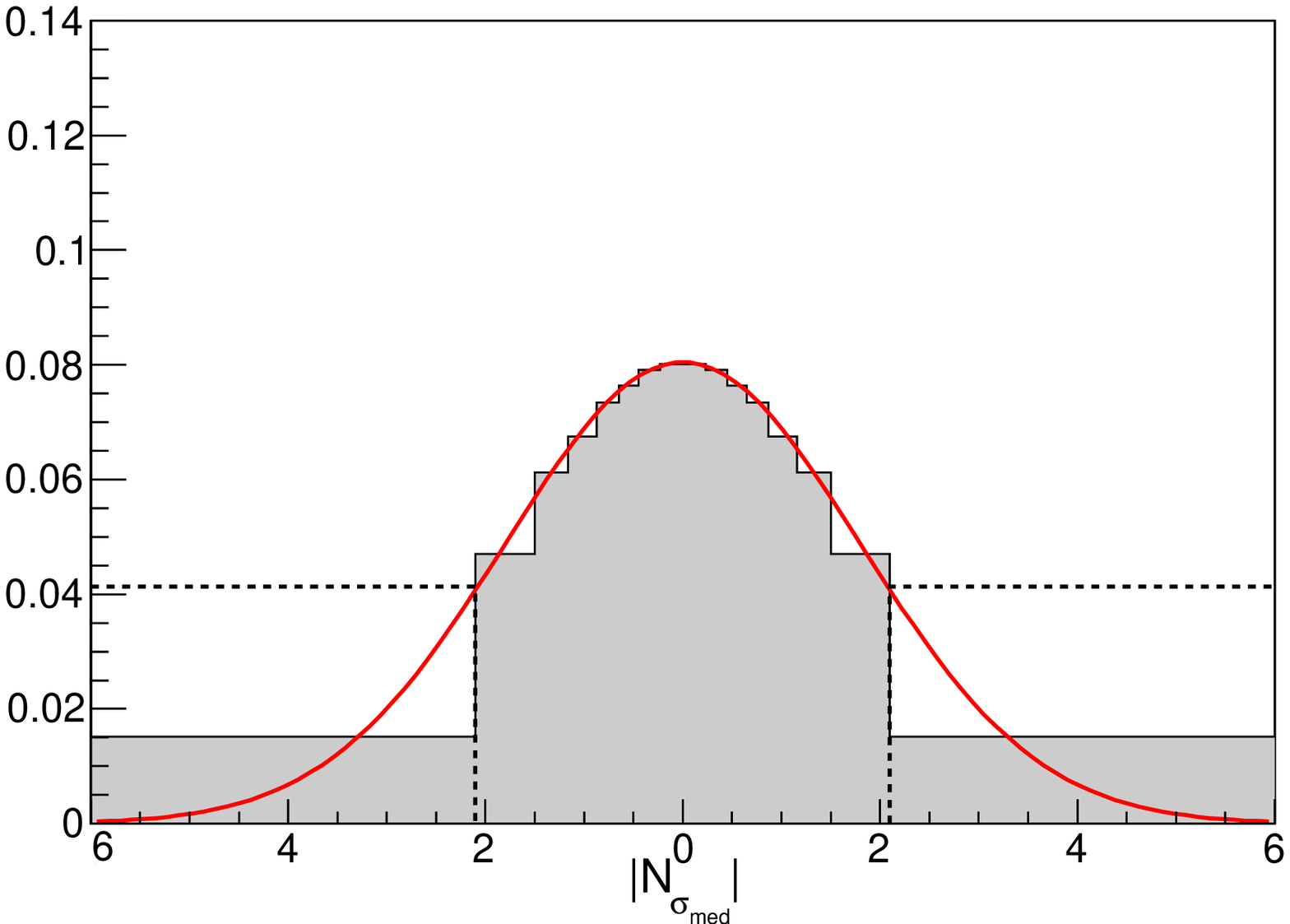}
\caption{Best fit Gaussian probability density functions. The top left (right) plot represents the $|N_{\sigma_{\mathrm{wm}}}|$ error distribution with scale factor $S=1$ (1.7). The bottom left (right) plot represents the $|N_{\sigma_{\mathrm{med}}}|$ error distribution with scale factor $S=1$ (1.8). The dotted lines represent the predicted probability of the last bins brought in from $|N_{\sigma}|=\infty$ to $|N_{\sigma}|=6.0$ with their heights adjusted to maintain the same probability.}  
\label{figure:Gauss}
\end{figure}
\end{center}

Although we have noted the non-Gaussianity of the error data, we will begin with the Gaussian probability distribution function where $|N_{\sigma}|=1$ is synonymous with 1 standard deviation. The Gaussian distribution is
\begin{equation}
P(|N_{\sigma}|)=\frac{1}{\sqrt{2\pi}}\exp(-|N_{\sigma}|^2/2).
\end{equation}
We also consider the function $P(|N_{\sigma}|/S)$ where we allow $S$ to vary over 0.1--3, in steps of 0.1, and determine the value of $S$ that minimizes $\chi^2$. When the scale factor, $S$, is not included (or equal to 1), we have $\nu = 7$ degrees of freedom. However, when $S$ (an additional free parameter) is allowed to vary we have only $\nu = 6$ degrees of freedom. After normalizing both fits to unit area, shown in Fig. \ref{figure:Gauss}, we see that the measurement error distributions are poorly fitted by a Gaussian distribution (also see Table \ref{table:Goodness-of-Fit}). Allowing the width of the Gaussian curve to vary to minimize $\chi^2$ favors $S=1.7$ for the weighted mean case. In this case 1 standard deviation is represented by $|N_{\sigma}|=1.7$. This again points to the fact that the error distribution for the 66 measurements under consideration is non-Gaussian.\footnote{Although the error distributions of the lithium abundance measurements are non-Gaussian, this does not necessarily imply that the measurement errors themselves are non-Gaussian. Instead, it perhaps tells us something about the observers' ability to correctly estimate systematic and statistical uncertainties.}   

\begin{center}
\begin{figure}[H]
\advance\leftskip-1.25cm
\advance\rightskip-1.25cm
\includegraphics[height=68mm,width=95mm]{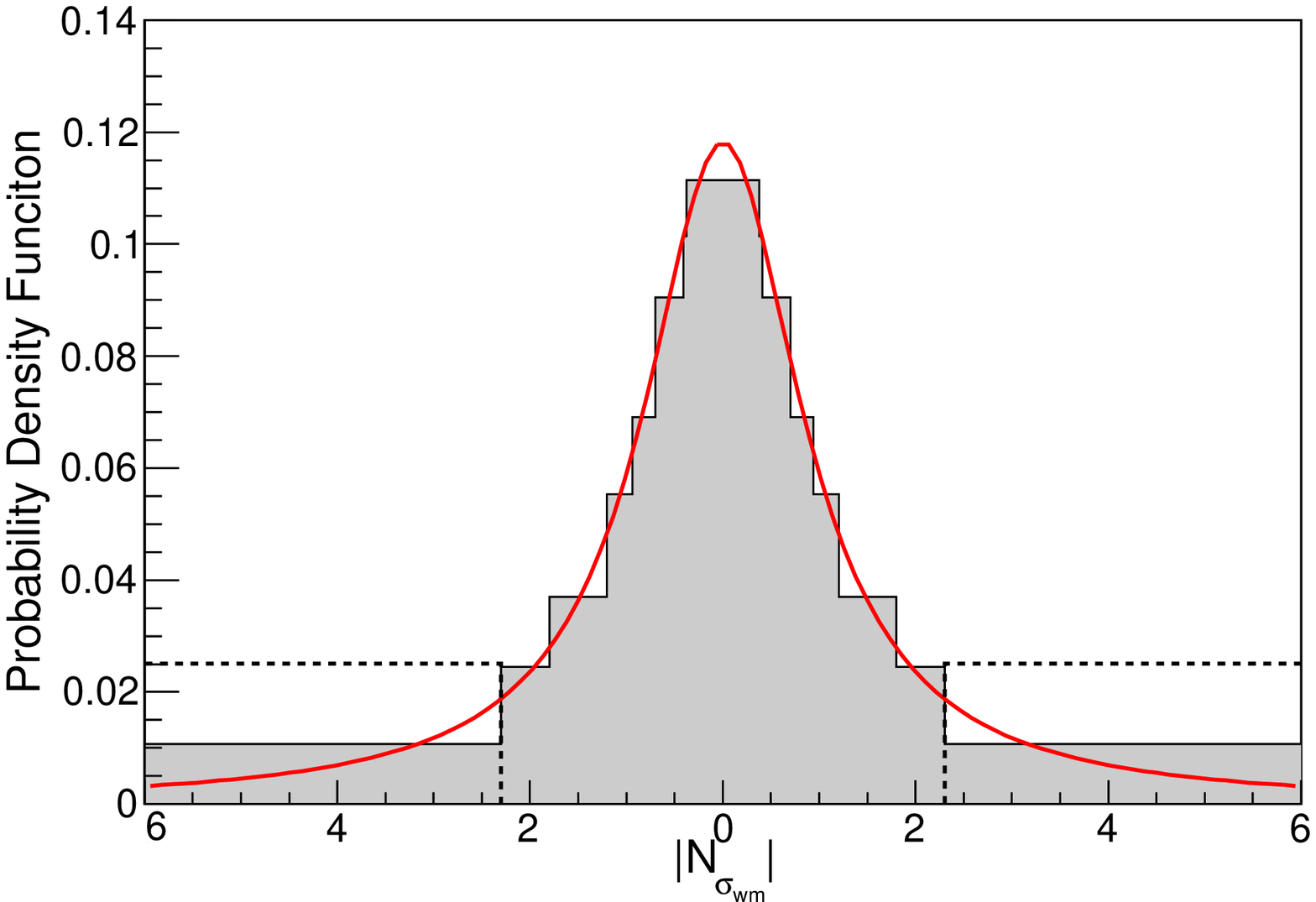}
\includegraphics[height=68mm,width=95mm]{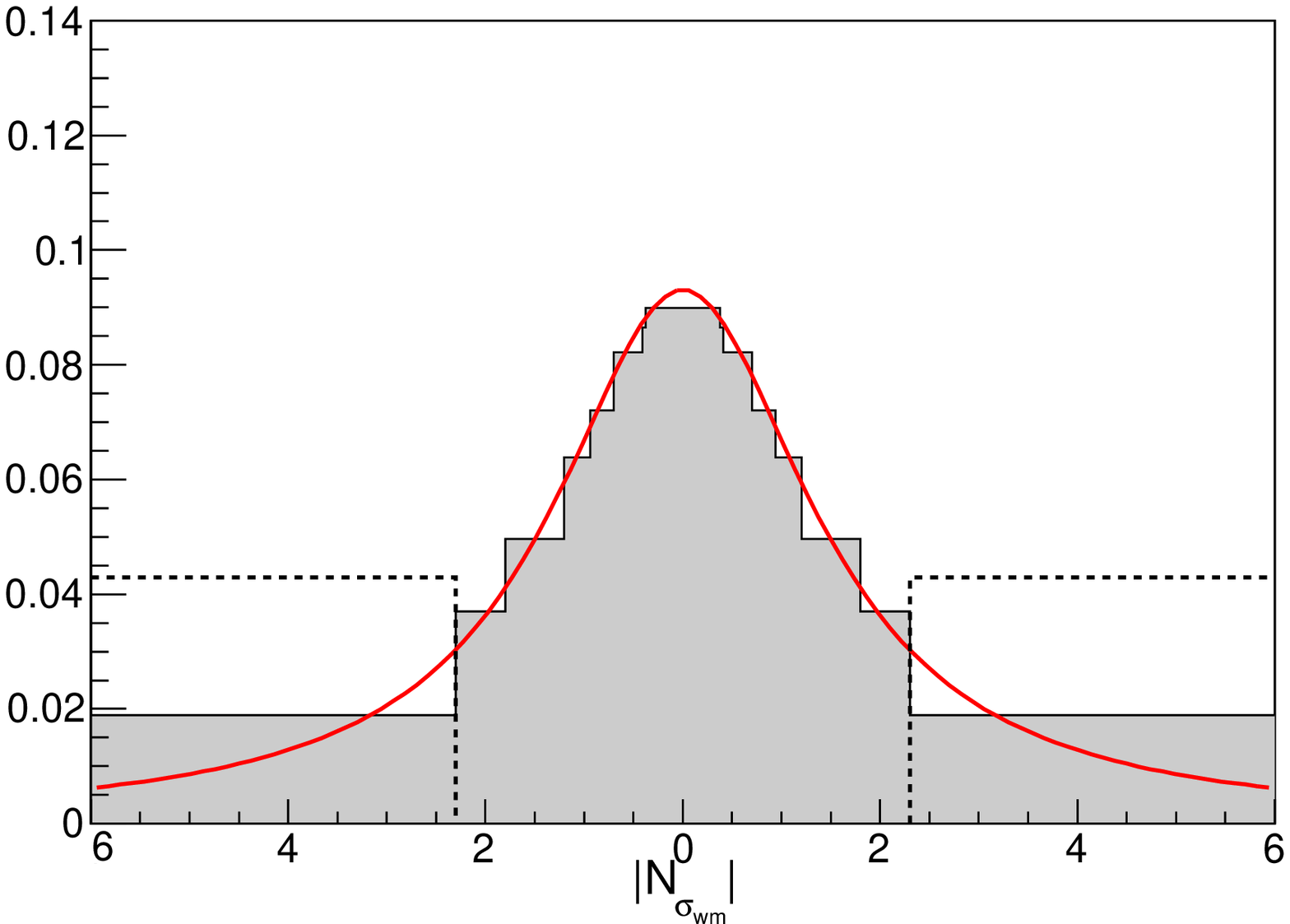}
\includegraphics[height=68mm,width=95mm]{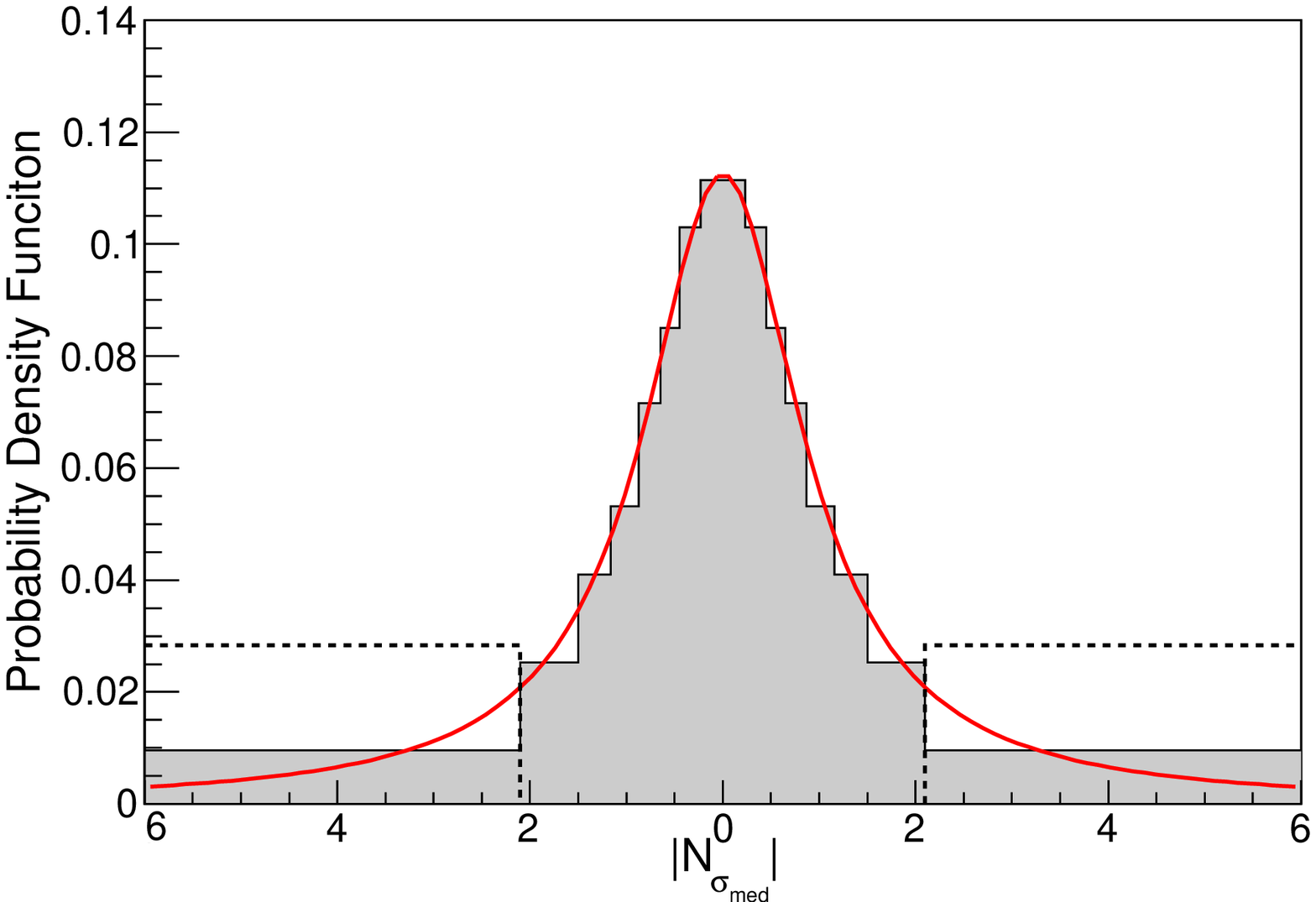}
\includegraphics[height=68mm,width=95mm]{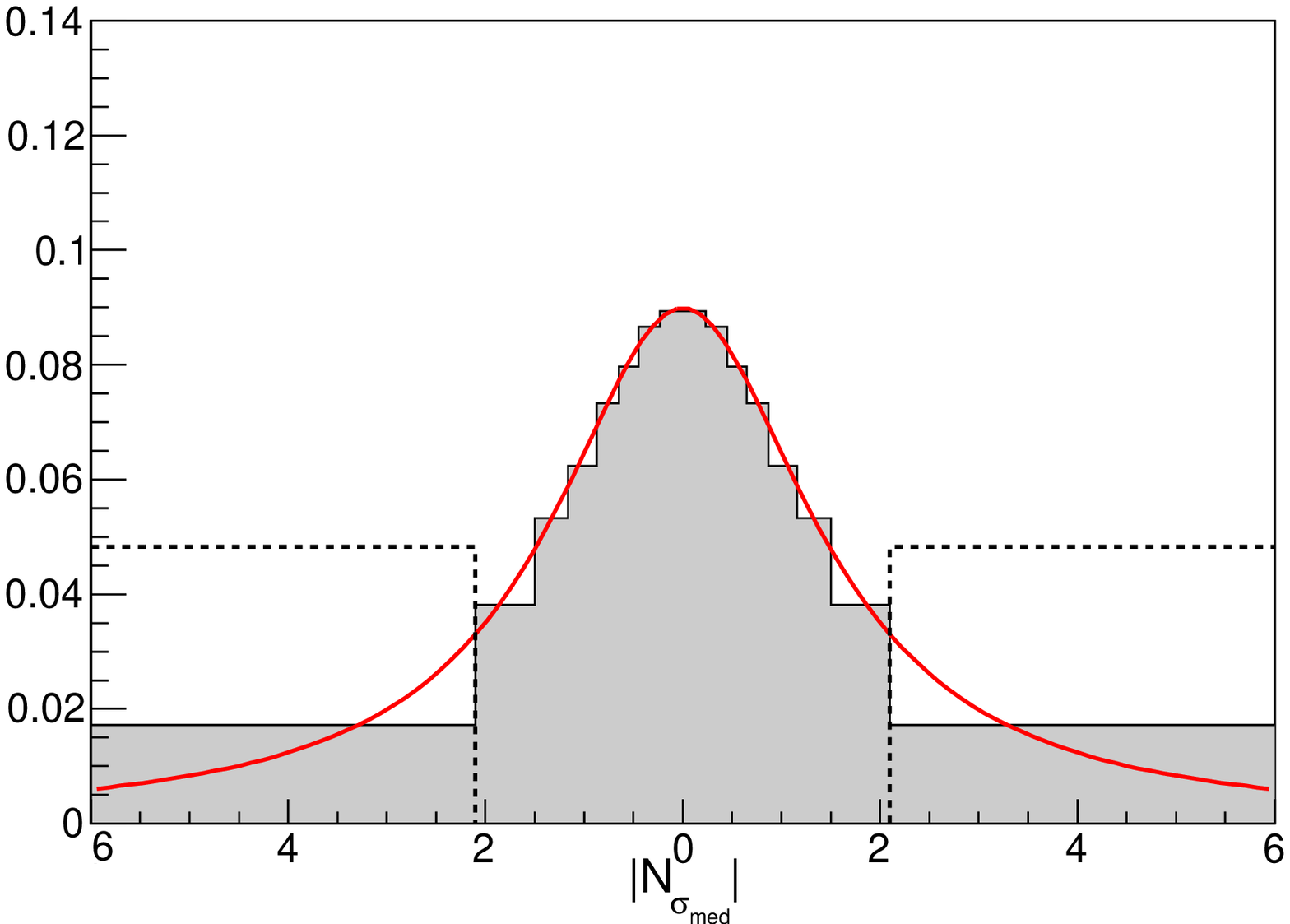}
\caption{Best fit Cauchy probability density functions. The top left (right) plot represents the $|N_{\sigma_{\mathrm{wm}}}|$ error distribution with scale factor $S=1$ (1.6). The bottom left (right) plot represents the $|N_{\sigma_{\mathrm{med}}}|$ error distribution with scale factor $S=1$ (1.6). The dotted lines represent the predicted probability of the last bins brought in from $|N_{\sigma}|=\infty$ to $|N_{\sigma}|=6.0$ with their heights adjusted to maintain the same probability.}  
\label{figure:Cauchy}
\end{figure}
\end{center}

Next, we turn to the Cauchy, or Lorentzian, distribution, which has an extended tail and is described by
\begin{equation}
P(|N_{\sigma}|)=\frac{1}{\pi}\frac{1}{1+|N_{\sigma}|^2}.
\end{equation}
Again, we also consider the case of $P(|N_{\sigma}|/S)$, where $S$ is the scale factor that is allowed to vary while $\chi^2$ is minimized. The best-fit Cauchy distributions for both weighted-mean and median central estimates are shown in Fig. \ref{figure:Cauchy}, and numerical values are listed in Table \ref{table:Goodness-of-Fit}. A Cauchy distribution with scale factor $S=1$ is a poor fit, similar to the Gaussian distribution case, with a probability less than 0.1\%. Although still not impressive, but a significant improvement nonetheless, the Cauchy distribution with a scale factor of $S=1.6$ has a significantly larger probability but is still less than 0.1\% although $\chi_{\nu}^2$ is cut nearly in half. Furthermore, as listed in Table \ref{table:Nsigma Limits}, it is apparent that the probability is much greater in the extended tails as 68.3\% and 95.4\% fall within $|N_{\sigma}| < 2.9$ and $|N_{\sigma}| < 22$, respectively, whereas the observed error distribution has the limits of $|N_{\sigma}| < 1.4$ and $|N_{\sigma}| < 3.0$. Therefore, it looks like a better fit will be one with broader tails than the Gaussian distribution but not as broad as that of the Cauchy distribution. 

\begin{deluxetable}{llcc}
\tablecaption{$|N_{\sigma}|$ Limits\tablenotemark{a}}
\tablewidth{0pt}
\tablehead{ 
\colhead{Function\tablenotemark}& 
\colhead{Scale\tablenotemark{b}}& 
\colhead{68.3$\%$\tablenotemark{c}}&
\colhead{95.4$\%$\tablenotemark{c}}\\
\noalign{\vskip -6mm}
}
\startdata
\noalign{\vskip -0mm}
\noalign{\vskip 1mm} 
Gaussian............................&	1&	1.0&	2.0\\
\noalign{\vskip 2mm}
Gaussian............................&	1.7&	1.7&	3.4\\
\noalign{\vskip 2mm}
Gaussian............................&	1.8&	1.8	&3.6\\
\noalign{\vskip 2mm}
Cauchy...............................&	1&	1.8&	14\\
\noalign{\vskip 2mm}
Cauchy...............................&	1.6&	2.9&	22\\
\noalign{\vskip 2mm}
Cauchy...............................&	1.6&2.9	&22	\\
\noalign{\vskip 2mm}
$n = 8$ Student's $t$...............&	1&	1.1&	 2.4\\
\noalign{\vskip 2mm}
$n = 8$ Student's $t$...............&	2.6&2.8&	6.1\\
\noalign{\vskip 2mm}
$n = 8$ Student's $t$...............&	2.8&	3.0&6.6	\\
\noalign{\vskip 2mm}
Double Exponential...........&	1&	1.2&	3.1\\
\noalign{\vskip 2mm}
Double Exponential...........&	1.4&1.6&	4.3\\
\noalign{\vskip 2mm}
Double Exponential...........&	1.5&	1.7&4.6\\
\noalign{\vskip 2mm}
Observed Weighted Mean&   & 1.4& 3.0\\
\noalign{\vskip 2mm}
Observed Median.............&   & 1.3& 3.1\\
\enddata
\label{table:Nsigma Limits}
\tablenotetext{a}{For each set of named distribution functions, the first line is for the standard distribution and the second and third lines are for the distributions that best fit the error distribution constructed using the weighted mean and median central estimate respectively.}
\tablenotetext{b}{For a Gaussian distribution, $N_{\sigma}=1$ corresponds to 1 standard deviation when the scale factor is $S=1$. For the other functions, unless $S=1$, the scale factor varies with the width of the distribution to allow $\chi^{2}$ to be minimized.}
\tablenotetext{c}{The $|N_{\sigma}|$ limits that contain 68.3\% and 95.4\% of the probability.}
\end{deluxetable}

This brings us to the consideration of the Student's $t$ distribution which is described by the equation
\begin{equation}
P_n(|N_{\sigma}|)=\frac{\Gamma[(n+1)/2]}{\sqrt{\pi{n}}\Gamma(n/2)}\frac{1}{(1+|N_{\sigma}|^2/n)^{(n+1)/2}},
\end{equation}
where $\Gamma$ is the gamma function and $n$ is a positive integer (which consequently eliminates another degree of freedom; so now when not varying the scale factor $\nu=6$). As usual, we also consider the case of $P_n(|N_{\sigma}|/S)$, where $S$ is allowed to vary in such a way (from 0.1--3) to allow $\chi^2$ to be minimized. We consider Student's $t$ distribution because at $n=1$ Student's $t$ distribution is synonymous with the Cauchy distribution, and as $n\rightarrow\infty$ Student's $t$ distribution approaches a Gaussian distribution. Therefore, Student's $t$ distribution will have broader tails than the Gaussian distribution and narrower tails than the Cauchy distribution when $1<n<\infty$, along the lines of what the A(Li) error distributions seem to demand.

\begin{center}
\begin{figure}[H]
\advance\leftskip-1.25cm
\advance\rightskip-1.25cm
\includegraphics[height=68mm,width=95mm]{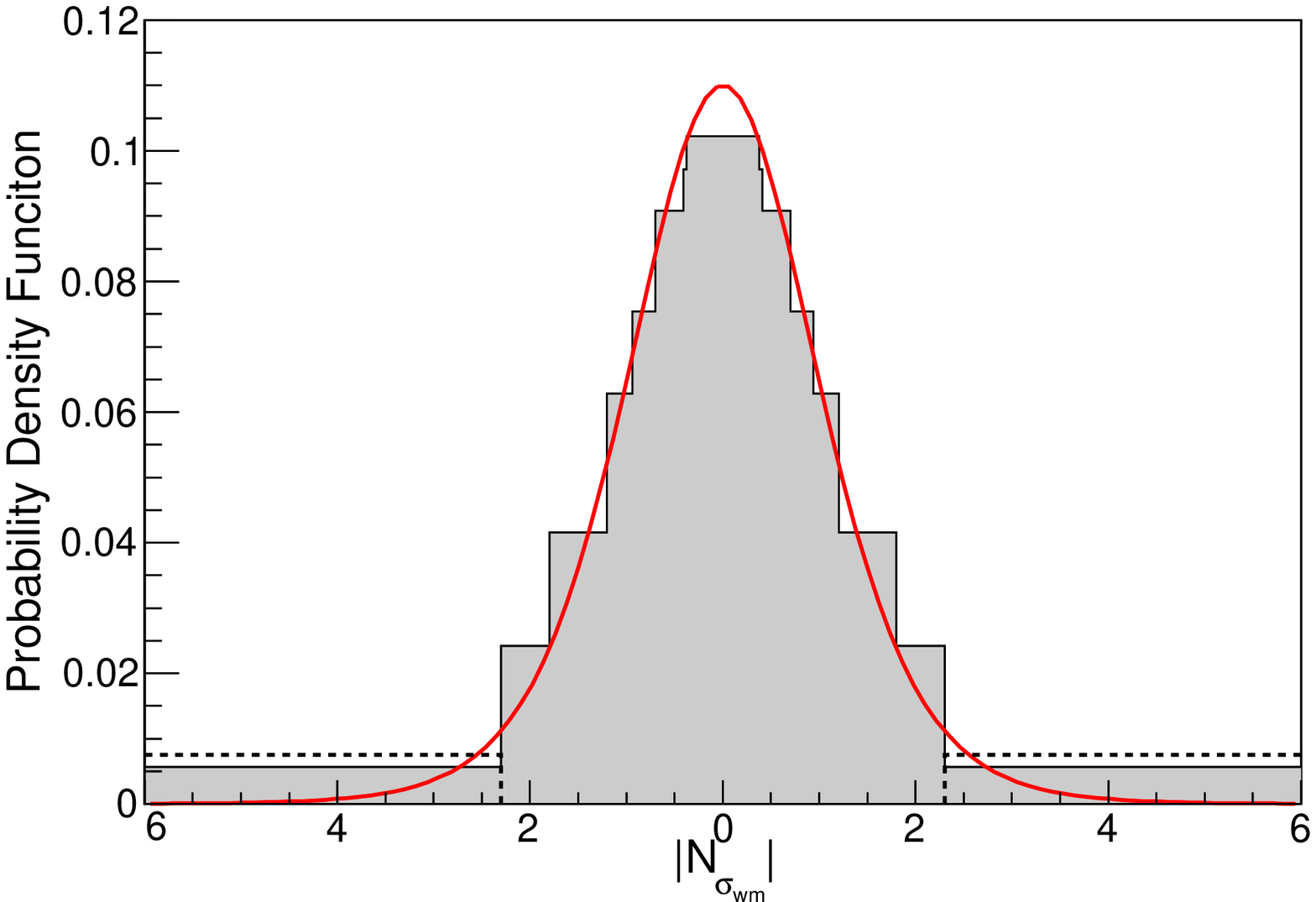}
\includegraphics[height=68mm,width=95mm]{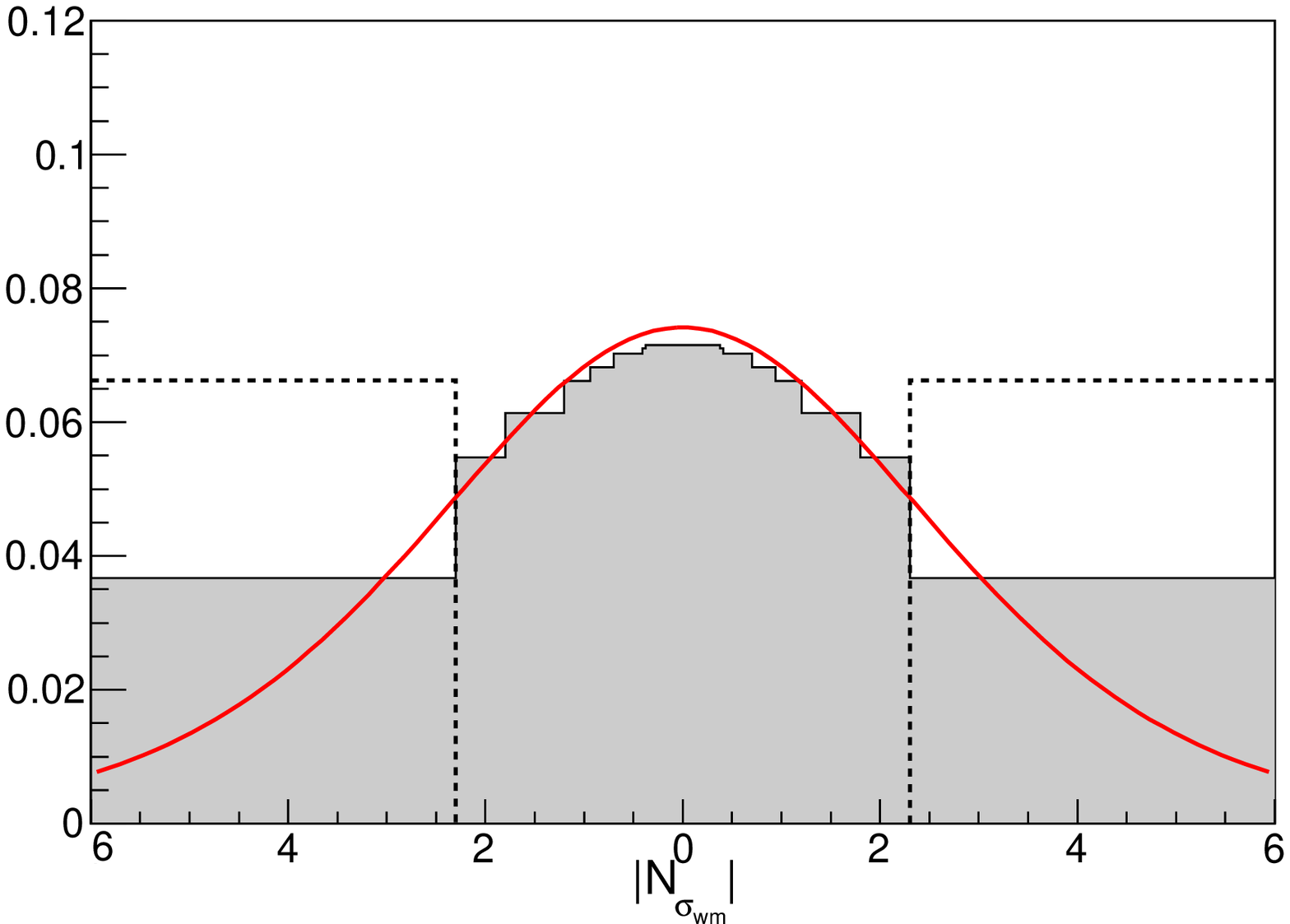}
\includegraphics[height=68mm,width=95mm]{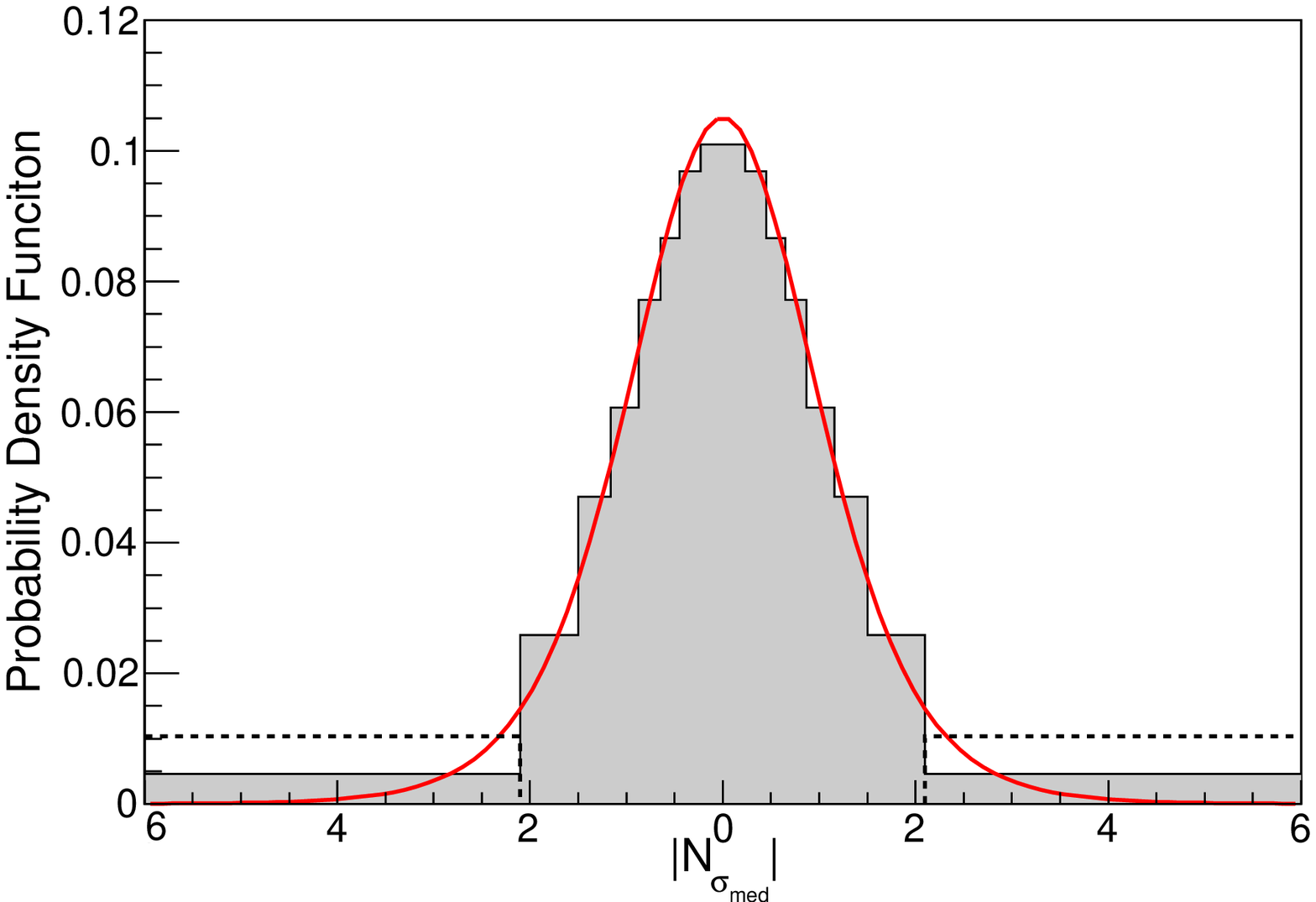}
\includegraphics[height=68mm,width=95mm]{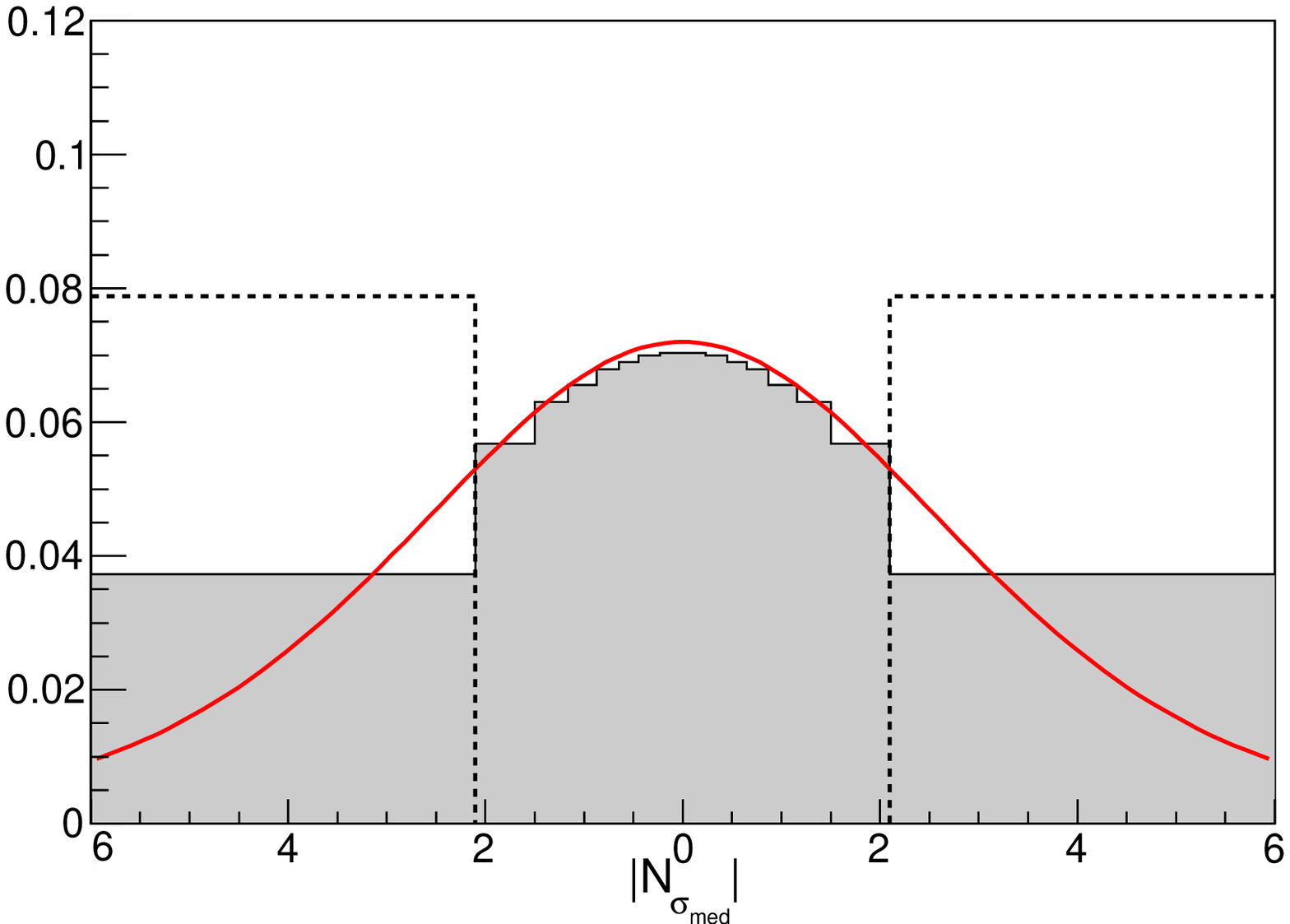}
\caption{Best fit Student's $t$ probability density functions. The top left (right) plot represents the $|N_{\sigma_{\mathrm{wm}}}|$ error distribution with scale factor $S=1$ (2.6) and $n=8$. The bottom left (right) plot represents the $|N_{\sigma_{\mathrm{med}}}|$ error distribution with scale factor $S=1$ (2.8) and $n=8$. The dotted lines represent the predicted probability of the last bins brought in from $|N_{\sigma}|=\infty$ to $|N_{\sigma}|=6.0$ with their heights adjusted to maintain the same probability.}  
\label{figure:student}
\end{figure}
\end{center}

We have fitted Student's $t$ distribution to the error distributions of A(Li) while allowing $n$ to vary over integer values between 2 and 30 as $\chi_{\nu}^2$ is minimized (see Fig. \ref{figure:student}). When considering the corresponding $P_n(|N_{\sigma}|/S)$ (in which case $\nu=5$), we  allow $n$ to be any integer value between 2 and 30 while simultaneously allowing $S$ to adjust over $0.1-3$ for the best $\chi_{\nu}^{2}$.\footnote{When allowing $n$ to vary, a pattern began to form that resulted in a progressively better fit for every even value of $n$. As we looked for the smallest $\chi_{\nu}^2$ for each value of $n$, $S$ had to be raised exponentially (we exceeded our upper bound of 3 on $S$). For $n=10$ and $S=3.1$ there was a reduced $\chi^2$ probability of $\sim29$\%, for $n=12$ and $S=3.8$ the probability was $\sim61$\%, for $n=14$ and $S=4.7$ the probability was $\sim86$\%, and for $n=20$, we must have approached an asymptote because for $S>1000$ $\chi_{\nu}^2$ showed very gradual change, while still minimizing, resulting in a probability of $\sim99.9$\%.} In this case the best fit occurs when $n=8$ and $S=2.6$\footnote{This, of course, only applies to our arbitrary limits of $0.1\leq S\leq3.0$, which we will consider our ``best fit" probability distribution from this point forward.} where $\chi^2$ is minimized, resulting in the highest probability of $6.9$\% (Table \ref{table:Goodness-of-Fit}). With Student's $t$ fit, only $29$\% of the data falls within $|N_{\sigma}|<1$ and $68.3$\% of the data falls within $|N_{\sigma}|<2.8$ as expressed in Table \ref{table:Nsigma Limits} and Table \ref{table:Expected Fraction}.

\begin{deluxetable}{llcc}
\tablecaption{Expected Fractions\tablenotemark{a}}
\tablewidth{0pt}
\tablehead{ 
\colhead{Function\tablenotemark}& 
\colhead{Scale\tablenotemark{b}}& 
\colhead{$|N_{\sigma}|\leq1$\tablenotemark{c}}&
\colhead{$|N_{\sigma}|\leq2$\tablenotemark{c}}\\
\noalign{\vskip -6mm}
}
\startdata
\noalign{\vskip -0mm}
\noalign{\vskip 1mm} 
Gaussian............................&	1&	0.68&	0.95\\
\noalign{\vskip 2mm}
Gaussian............................&	1.7&	0.44&	0.76\\
\noalign{\vskip 2mm}
Gaussian............................&	1.8&	0.42&	0.73\\
\noalign{\vskip 2mm}
Cauchy...............................&	1&	0.50&	0.71\\
\noalign{\vskip 2mm}
Cauchy...............................&	1.6&	0.36&	0.57\\
\noalign{\vskip 2mm}
Cauchy...............................&	1.6&0.36	&	0.57	\\
\noalign{\vskip 2mm}
$n = 8$ Student's $t$...............&	1&	0.65&	0.92\\
\noalign{\vskip 2mm}
$n = 8$ Student's $t$...............&	2.6&0.29&	0.54\\
\noalign{\vskip 2mm}
$n = 8$ Student's $t$...............&	2.8&0.27	&	0.50	\\
\noalign{\vskip 2mm}
Double Exponential...........&	1&	0.63&	0.87\\
\noalign{\vskip 2mm}
Double Exponential...........&	1.4&0.51&	0.76\\
\noalign{\vskip 2mm}
Double Exponential...........&	1.5&	0.49&	0.74	\\
\noalign{\vskip 2mm}
Observed Weighted Mean&   & 0.55&0.82\\
\noalign{\vskip 2mm}
Observed Median.............&   & 0.52& 0.82\\
\enddata
\label{table:Expected Fraction}
\tablenotetext{a}{For each set of named distribution functions, the first line is for the standard distribution and the second and third lines are for the distributions that best fit the error distribution constructed using the weighted mean and median central estimate respectively.}
\tablenotetext{b}{For a Gaussian distribution, $N_{\sigma}=1$ corresponds to 1 standard deviation when the scale factor is $S=1$. For the other functions, unless $S=1$, the scale factor varies with the width of the distribution to allow $\chi^{2}$ to be minimized.}
\tablenotetext{c}{The fraction of the area that lies within $|N_{\sigma}|=1$ and $|N_{\sigma}|=2$.}
\end{deluxetable}

\begin{center}
\begin{figure}[H]
\advance\leftskip-1.25cm
\advance\rightskip-1.25cm
\includegraphics[height=68mm,width=95mm]{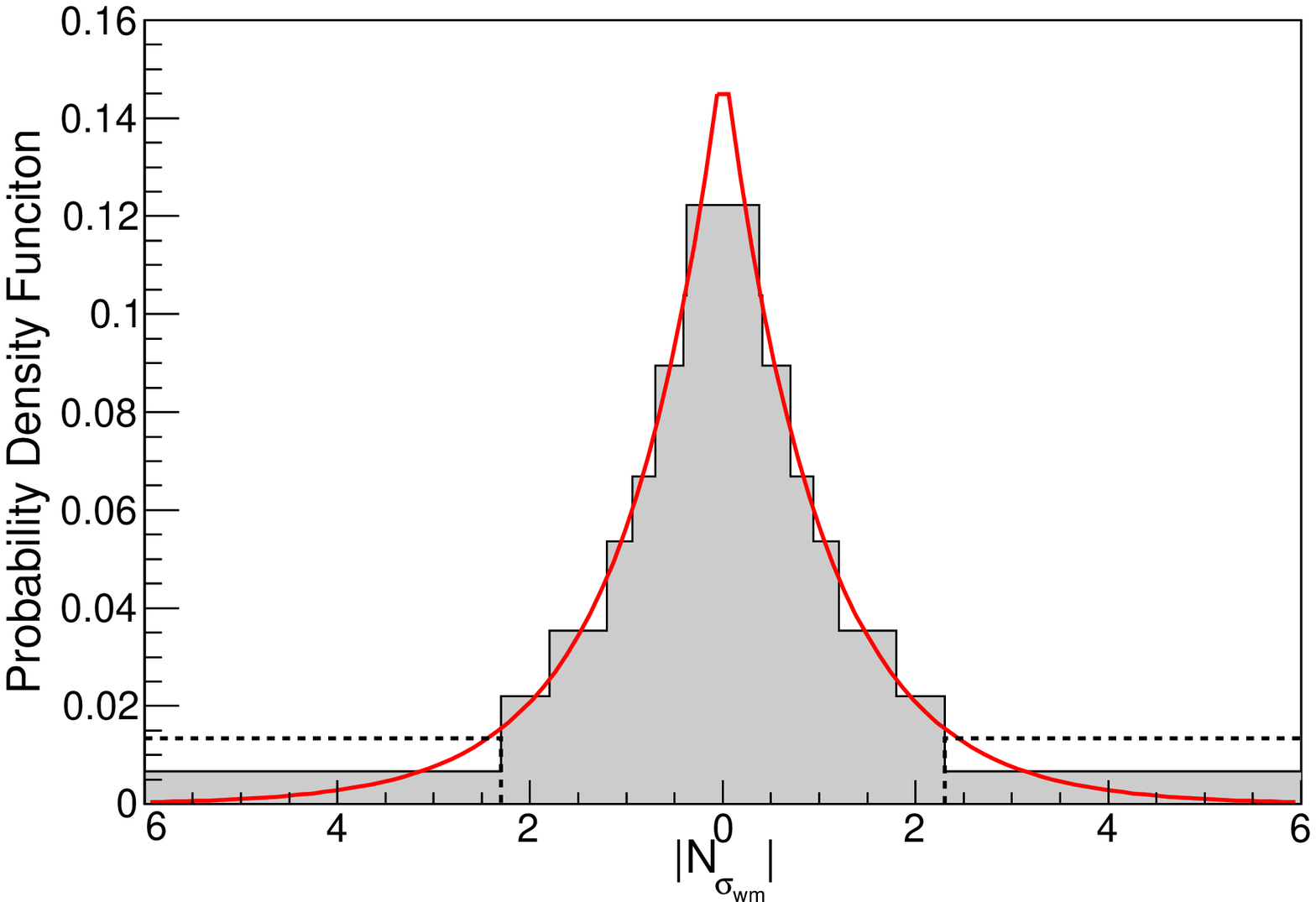}
\includegraphics[height=68mm,width=95mm]{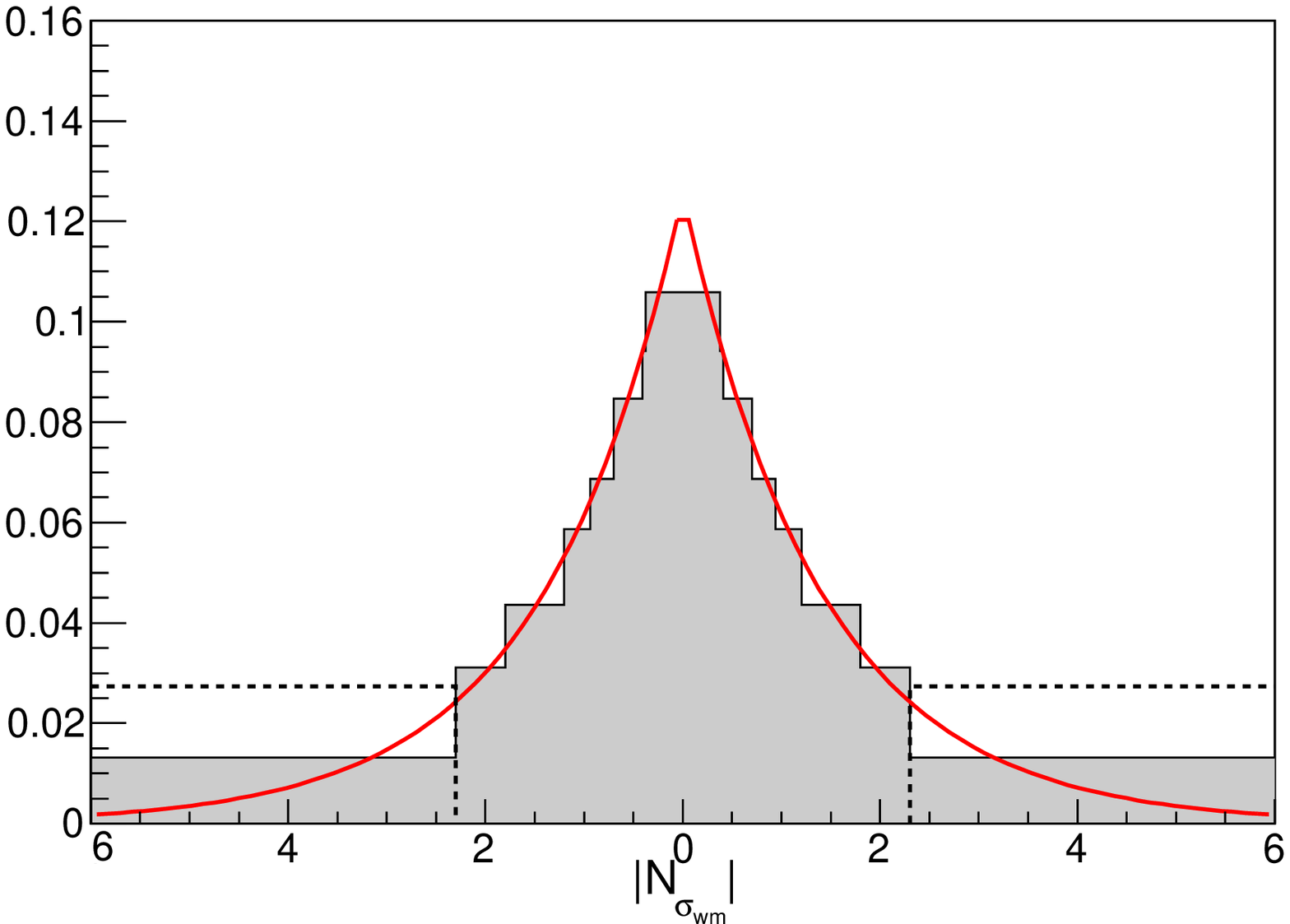}
\includegraphics[height=68mm,width=95mm]{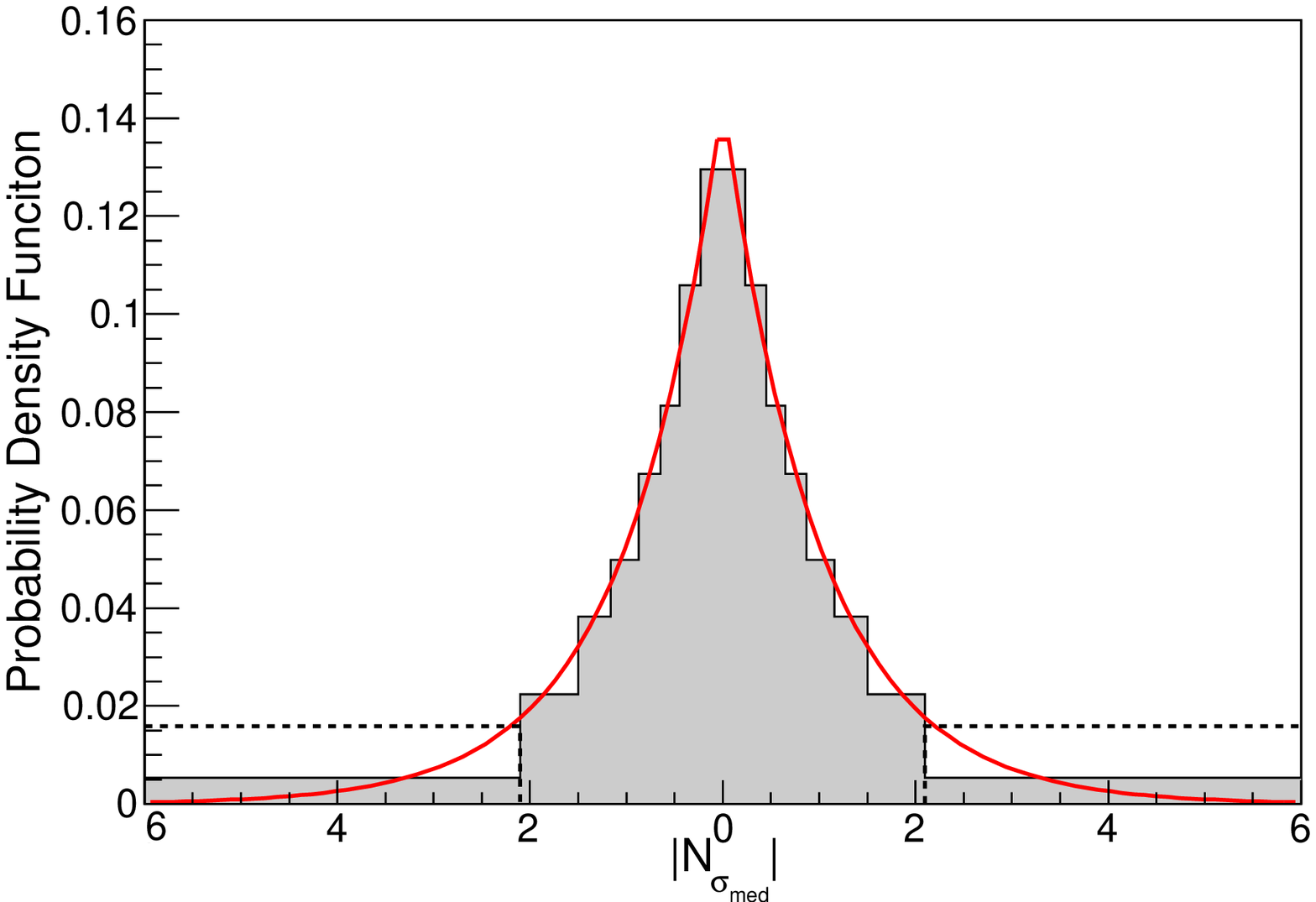}
\includegraphics[height=68mm,width=95mm]{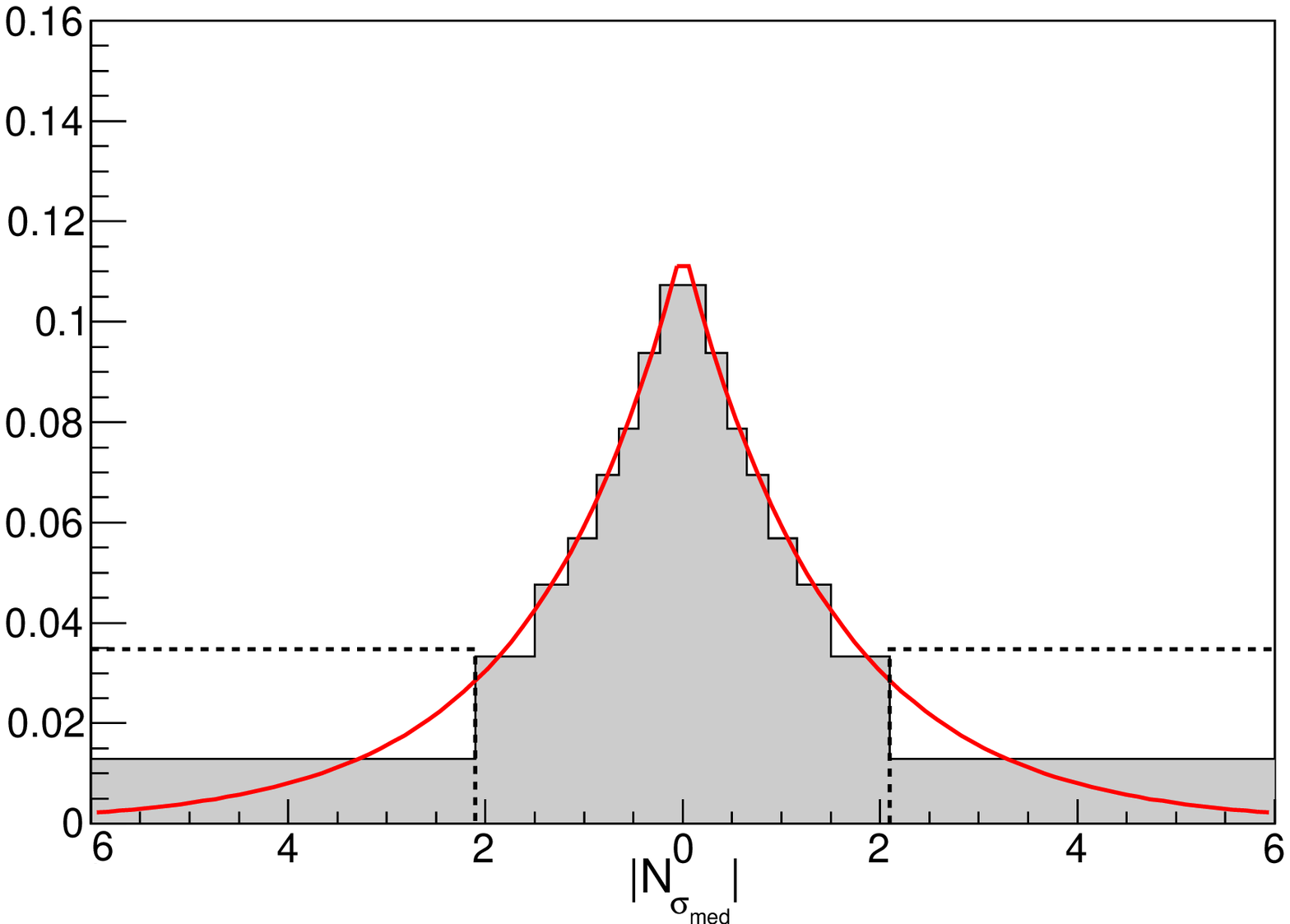}
\caption{Best fit double exponential probability density functions. The top left (right) plot represents the $|N_{\sigma_{\mathrm{wm}}}|$ error distribution with scale factor $S=1$ (1.4). The bottom left plot represents the $|N_{\sigma_{\mathrm{med}}}|$ error distribution with scale factor $S=1$ (1.5). The dotted lines represent the predicted probability of the last bins brought in from $|N_{\sigma}|=\infty$ to $|N_{\sigma}|=6.0$ with their heights adjusted to maintain the same probability.}  
\label{figure:DE}
\end{figure}
\end{center}

Finally, we consider the double exponential, or Laplace, distribution,
\begin{equation}
P(|N_{\sigma}|)=\frac{1}{2}e^{-|N_{\sigma}|}.
\end{equation}
The double exponential distribution falls off more rapidly than the Cauchy distribution, but not as quickly as the Gaussian distribution. When considering $P_n(|N_{\sigma}|/S)$ we found the best fit was when $S=1.4$, resulting in $68.3$\% of the data falling within $|N_{\sigma}|=1.6$ and $95.4$\% of the data within $|N_{\sigma}|=4.3$ which is shown numerically in Table \ref{table:Nsigma Limits} and visually in Figure \ref{figure:DE}.

\section{Conclusion}
\label{Conclusion}
We have used a compilation of 66 $^{7}\rm{Li}$ abundance measurements from \cite{Spite12} to attempt to gain a better understanding of the lithium problem. Confirming the observation of \cite{Spite10}, we find that the A(Li) error distribution is non-Gaussian. As noted above, this perhaps tells us something about the observers' ability to estimate systematic and statistical uncertainties.\footnote{We examine an illustrative example in the Appendix.} This could also be an indication that a definite resolution might need to wait for more and higher-quality data. However, here we speculate a bit about the statistical significance of the Li problem.  

To determine the statistical significance of the discrepancy, we first assume Gaussianity and find the difference between the mean of the $Planck$ data predicted value $\rm{A(Li)}=2.69$ \citep{Coc2014} and our median value $\rm{A(Li)}=2.21$. We divide this difference by the quadrature sum of our error ($\sigma=0.065$) and that from \cite{Coc2014} ($\sigma=0.034$). This results in a $6.5\sigma$ discrepancy. To account for the non-Gaussianity, we simply multiply our $\sigma=0.065$ error by 1.4 (Table \ref{table:Nsigma Limits}, second to last line) in the quadrature sum, resulting in a $4.9\sigma$ discrepancy. 

To attempt to characterize the A(Li) error distribution, we fit various popular probability distribution functions to it. We find that the observed error distribution has larger probability in the tails than a Gaussian distribution, but less than a Cauchy distribution. While allowing the scale factor to vary over the range $0.1{\leq}S{\leq}3$, the error distribution is best fit by an $n=8$ Student's $t$ distribution, although better fits can be found for larger $n$ and $S$. We have not followed up on this in detail since this is unlikely to be of much physical significance.

In conclusion, while it would be good to have more and higher-quality A(Li) data that results in a Gaussian error distribution so as to be able to draw a definite conclusion, it seems fair to conclude that the non-Gaussianity of the current data cannot fully resolve the Li problem.

We thank M. Spite for providing us with helpful insight and the data compiled in \cite{Spite12}. We are grateful to F. Spite for very useful advice. This work was supported in part by DOE Grant No. DEFG 03-99EP41093, and NSF Grant Nos. AST-1109275 and PHY-1157044.

\begin{appendices}
\section*{Appendix}

\begin{center}
\begin{figure}[H]
\advance\leftskip-1.25cm
\advance\rightskip-1.25cm
\includegraphics[height=68mm,width=95mm]{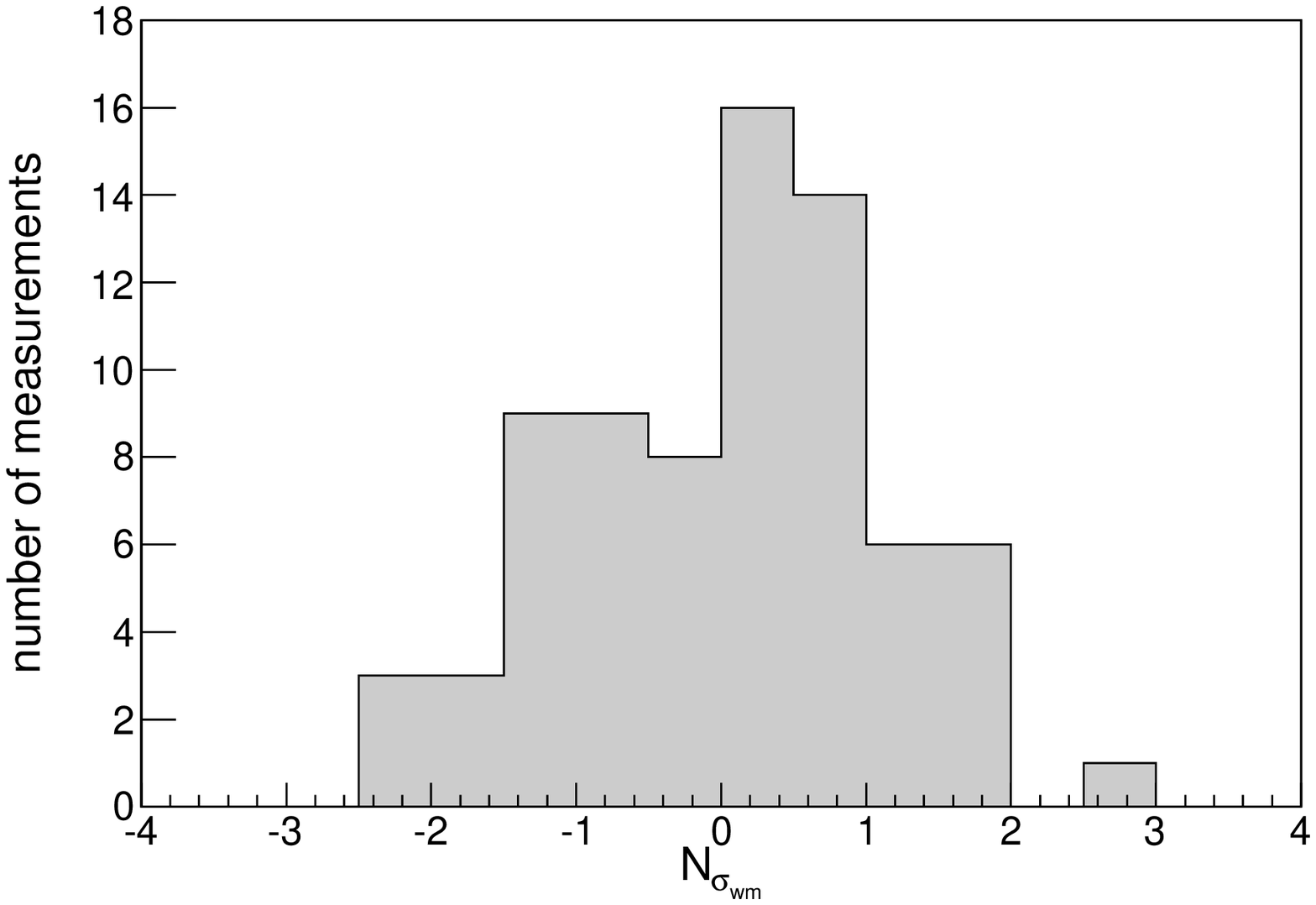}
\includegraphics[height=68mm,width=95mm]{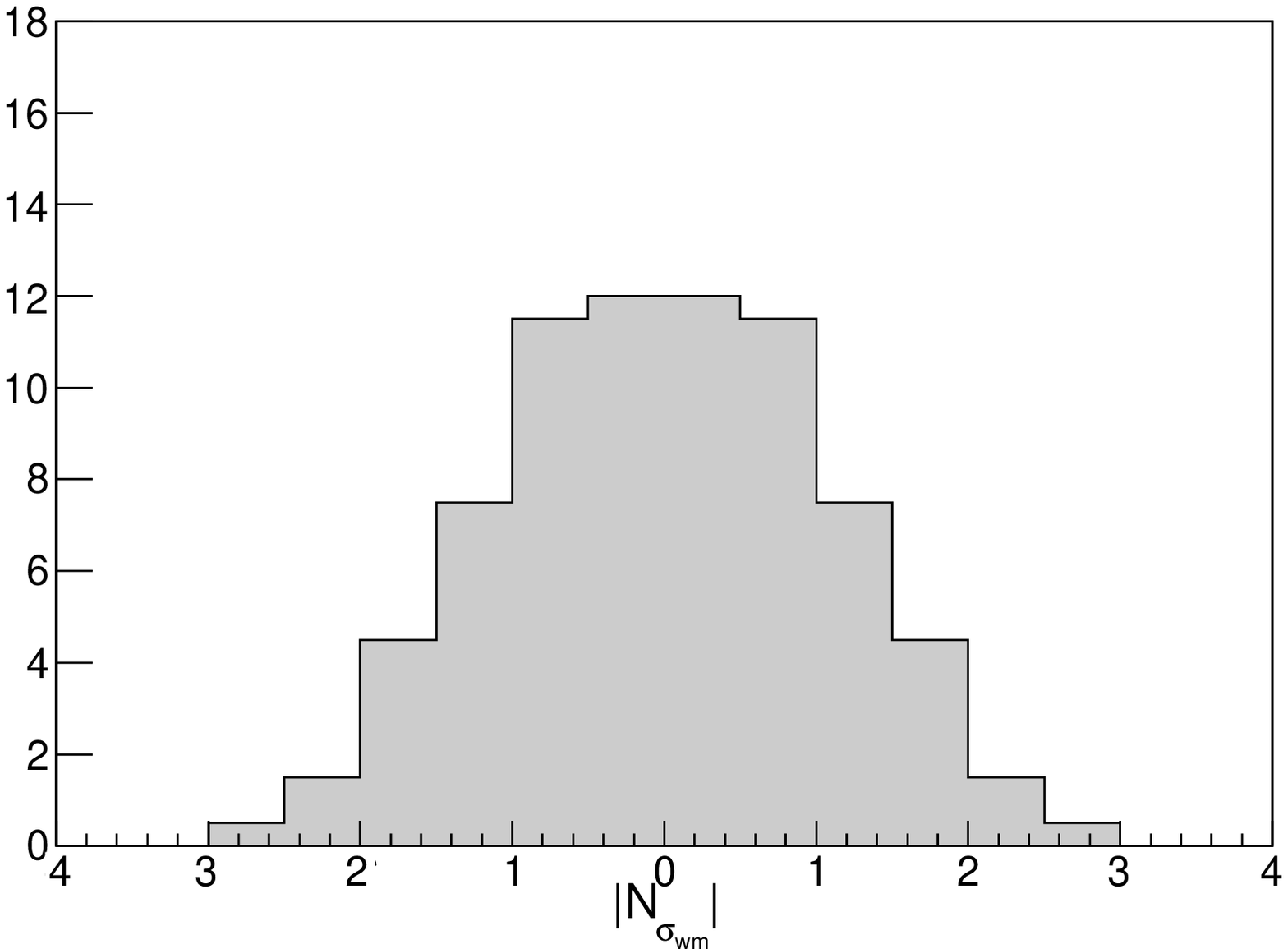}
\includegraphics[height=68mm,width=95mm]{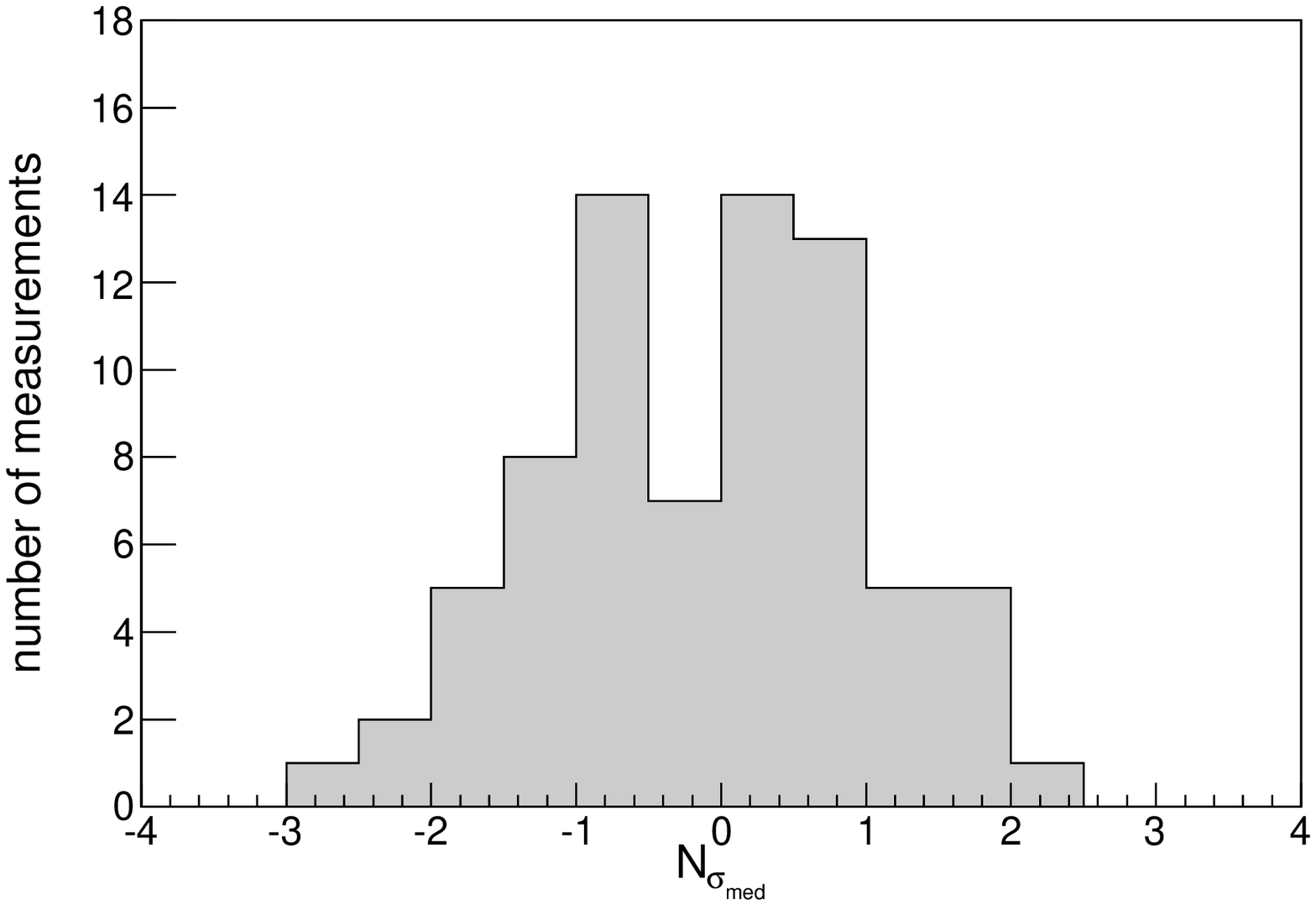}
\includegraphics[height=68mm,width=95mm]{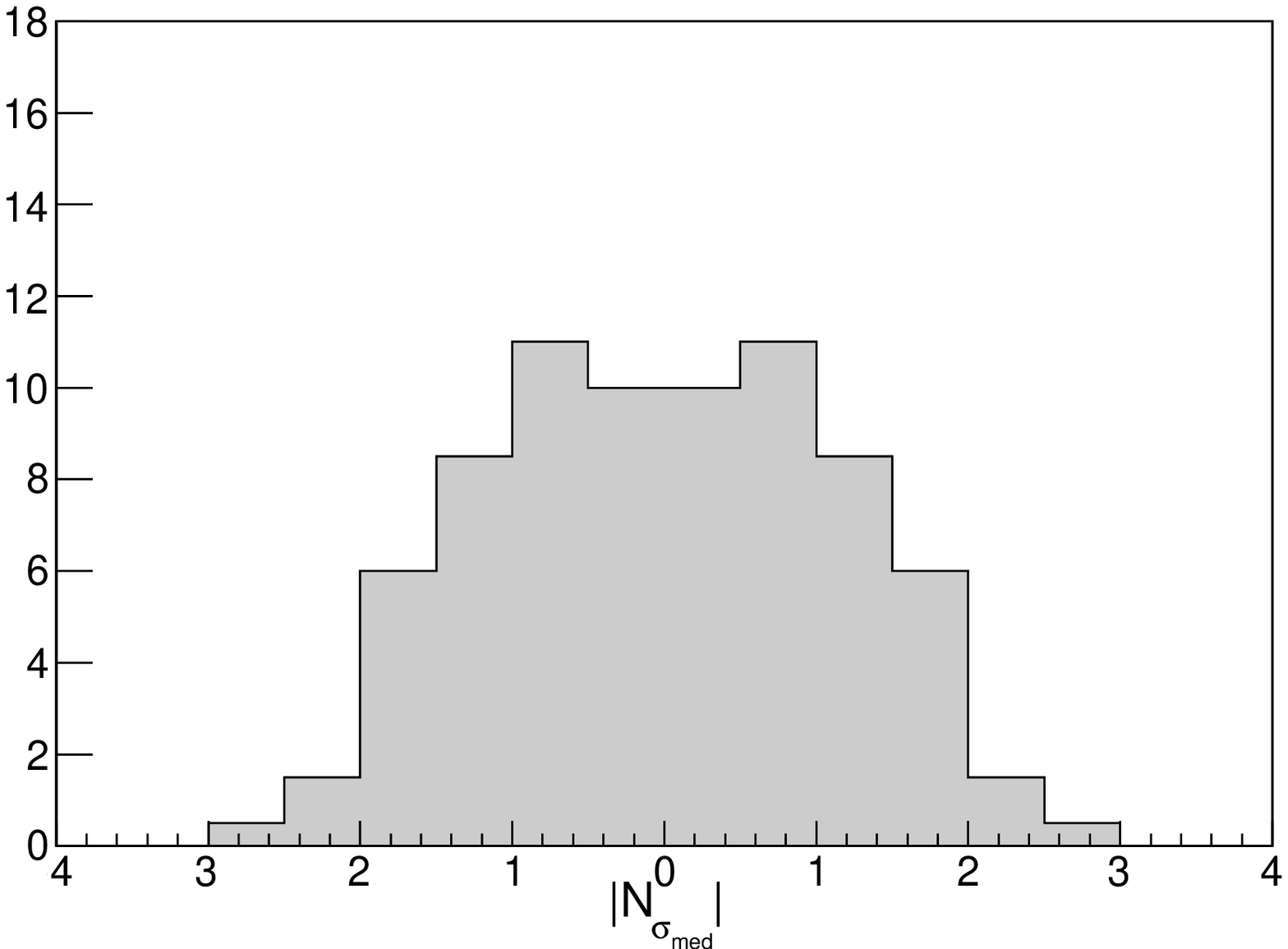}
\caption{Histograms of the error distribution with $\sigma=0.06$ A(Li) values in half standard deviation bins. The top (bottom) row uses the weighted mean (median) of the 66 measurements as the central estimate. The left (right) column shows the signed (absolute) deviation. In the left column plots, positive (negative) $N_{\sigma}$ represent a value that is greater (less) than the central estimate.}  
\label{figure:A}
\end{figure}
\end{center}

\cite{Bon07} have argued that the effective temperature error is larger than previously thought. If this is so then it could dominate the error, resulting in the same error for each A(Li) measurement. To illustrate the effect of such a potential error, we repeat our analysis with a constant, optimistic, $\sigma=0.06$.\footnote{We also used the more conservative $\sigma=0.09$ of \cite{Bon07} and reached similar conclusions.} For the weighted mean case 68.3\% of the signed error distribution falls within $-1.48\leq N_{\sigma}\leq0.80$ while 95.4\% lies in the range of $-2.47\leq N_{\sigma}\leq1.39$ and the absolute magnitude of the error distribution have corresponding limits of $|N_{\sigma}|\leq 1.15$ and $|N_{\sigma}|\leq 2.13$ respectively. For the median statistics central estimate 68.3\% of the signed error distribution falls within $-1.37\leq N_{\sigma}\leq0.83$ while 95.4\% lies within $-2.39\leq N_{\sigma}\leq1.60$ and the corresponding absolute magnitude limits are $|N_{\sigma}|\leq 1.16$ and $|N_{\sigma}|\leq 2.32$ respectively. Alternatively, when looking at the fraction of the data that falls within the $|N_{\sigma}| = 1$ and $2$ ranges respectively, we obtain $62.7\%$ and $94.7\%$ for the weighted mean case and $65.3\%$ and $95.4\%$ for the median one.  As one might expect, these fractions are more Gaussian than values found in the analysis in the body of our paper. The $N_{\sigma}$ histograms are shown in Fig. \ref{figure:A}. 

\end{appendices}


\def\mnras{MNRAS}
\def\aapr{A\&A~Rev.}
\def\jcap{J. Cosmology Astropart. Phys.}
\def\apjl{ApJ}
\def\pasp{PASP}
\def\aap{A\&A}
\def\apss{Ap\&SS}
\def\apjs{ApJS}

\end{document}